# Ballistic propagation-mediated super-resolution light-element imaging: in-situ tracking dynamic hydrogen migration

Zhiyao Liang,[1,7] Feng Liu,[1,7] Linyuan Chen,[1,7] Zhiwei Lv,[1] Shiya Chen,[2] Xinyu Liao,[1] Yuzhou He,[3] Yi Wang,[1] Pengpeng Liu,[1] Hua Bing Tao,[1] Nianpeng Lu,[3] Changlin Zheng,[4] Yang Sun,[2] Qinghua Zhang,[3] Rafal E. Dunin-Borkowski,[5] Xian-Kui Wei[1,5,6]*

[1]College of Chemistry and Chemical Engineering, Xiamen University, 361005, Xiamen, China

[2]Department of Physics, Xiamen University, Xiamen, 361005, China

[3]Beijing National Laboratory for Condensed Matter Physics, Institute of Physics, Chinese Academy of Science, Beijing 100190, China

[4]State Key Laboratory of Surface Physics and Department of Physics, Fudan University, Shanghai, 200433, China

[5]Ernst Ruska-Centre for Microscopy and Spectroscopy with Electrons, Forschungszentrum Jülich, Jülich, 52428, Germany

[6]Innovation Laboratory for Sciences and Technologies of Energy Materials of Fujian Province (IKKEM), Xiamen 361005, China

[7]These authors contribute equally: Zhiyao Liang, Feng Liu, Linyuan Chen

Corresponding author Email: xkwei@xmu.edu.cn

**Abstract**

**Transmission electrons – with few-picometer wavelengths – can theoretically resolve all atoms in solids. However, constrained by factors such as lens aberrations, Z-contrast dependence and poor temporal resolution, the scanning transmission electron microscopy (STEM) fails to tackle the long-standing challenge of hydrogen dynamics on atomic scale. Here, our experiment-guided simulations reveal that unlike heavy-atom imaging - governed by wave-based imaging process - ballistic propagation of particle-like electrons in Cs-corrected TEM enables super-resolution imaging (SRI) of light atoms, achieving a resolution down to 14 pm and a robust hydrogen contrast of ~24%. Our in-situ SRI study not only uncovers polymetric hydrogen aggregation in $TiH_x$ and H-$LaAlO_3$, but also inter-site migration, rotational displacement of hydrogen species and collective hydrogen tunneling-induced polarization surge in the oxide. By exploiting the novel electron-matter interactions, our findings offer an unparalleled approach to probe hydrogen dynamics and establish accurate structure-property relationships in diverse functional materials.**

Transmission electron microscopy (TEM) provides unprecedented spatiotemporal resolution and plays critical roles in advancing the study of micro-to-nanoscale structures in the fields of physics, chemistry, energy, materials and engineering sciences. Its characterization capability is enhanced by utilization of state-of-the-art aberration-corrected optics, methodology development, in-situ external-field stimulation, and operando reaction monitoring in gaseous and liquid environments. Benefiting from this, remarkable research progress achieved in recent years includes atomic-level 3D nanostructure determination via atomic electron tomography[1], the attainment of 14 pm spatial resolution by virtue of electron ptychography[2, 3], single Pt-atom tracking in a liquid environment by high-angle annular dark-field imaging[4], atomic imaging of zeolite-confined single molecules using integrated differential phase contrast[5]. Among these techniques, atomic number Z-contrast scanning transmission electron microscopy (STEM) serves as a powerful platform to resolve structural motifs and establish structure-property relationships.

To attain superior material functionality, probing the migration dynamics of light-element ions in heavy-atom frameworks is critical for uncovering the real-time microscopic mechanisms behind energy storage, conversion and performance evolution. This is exemplified by charge-mediated transport of Li, H and O species in batteries and fuel cells[6, 7, 8, 9]. Nevertheless, this practical requirement precisely exposes the limitation and inherent bottleneck of STEM imaging. The main reason is that the contrast of light elements is severely suppressed by heavy elements, which arises from the Z-dependent scattering power of atoms to the incident electrons. This is exemplified by the weak and diffuse hydrogen contrast observed in metal hydrides and oxides[10, 11, 12], e.g., ~ 4% in $PdH_x$[13]. The intrinsic contrast limitation is further compounded by poor temporal resolution and low image signal-to-noise ratio. Consequently, fundamental questions regarding atomic-scale hydrogen dynamics remain unresolved and enigmas tied to hydrogen-induced performance modulation, e.g., hydrogen spillover in catalysis and embrittlement in metals[14, 15], persist as unsolved black-box issues.

Distinct from STEM, single-shot TEM imaging provides natural advantages for high temporal resolution[16, 17, 18]. Nevertheless, its capability of resolving hydrogen has not yet been validated. Apart from the weak beam electron-hydrogen interaction, another central challenge is resolving neighboring hydrogen columns with a projected separation of ≤ 70 pm, whereas the instrumental resolution itself is merely 70 pm. Therefore, overcoming the resolution barrier is essential to unravel the dynamics of hydrogen species. Over the past decades, the microscopy community has grappled with an issue termed as resolution dilemma, i.e., the Young's fringe resolution ($d_{YF}$) is persistently poorer than the information limit ($d_{info}$)[19], e.g., 70 pm vs ~50 pm at 300 kV[20], despite being measured under identical conditions. Another counterintuitive phenomenon is that atomic spacings larger than either value are not always resolvable, even for specimen only a few nanometers thick[21]. This inconsistency indicates an incomplete understanding of the fundamental electron-matter interaction principles.

To tackle these conundrums, we begin this work by quantifying structural motifs of $DyScO_3$ structural images acquired with a Cs-corrected transmission electron microscope (Cs-TEM). Unexpectedly, we found that the resolvable O-O splitting, rather than the larger Dy-Dy spacing, accounts for the experimental $d_{info}$. Guided by this insight, we carried out comparative experiments on $TiH_x$, H-$LaAlO_3$ and their H-free counterparts, which confirm the pivotal contribution of light atoms in enhancing the information limit and the $d_{info} < d_{YF}$ relationship. In contrast to the conventional wave-based description of heavy-atom contrast, our study clarifies that the weak Coulomb attraction from light atoms accentuates the particle character of electrons, and their ballistic propagation along light-atom columns enables the SRI. Leveraging the robust image quality and H-atom contrast, our in-situ SRI study unveils inter-site migration, rotational displacement of hydrogen, and collective hydrogen tunneling-induced polarization surge in H-$LaAlO_3$. The present work offers a powerful technical approach to explore field-driven dynamics of hydrogen in diverse functional materials.

## Limited- and super-resolution imaging of heavy atoms and oxygen

The orthorhombic $DyScO_3$ (space group Pbnm)[22] is characteristic of Dy-Dy splitting by $S_{Dy}$ = 70 pm and antiphase octahedral rotation ($a^-a^-a^+$) (Fig. 1a). In our experiments, a Cs-TEM with $d_{YF}$ = 70 pm is used to record atomic-resolution structural images under the negative Cs imaging condition[23] (Fig. 1b,c). Intriguingly, a nanoscale thickness gradient-induced paraelectric-to-ferroelectric (FE) phase transition is confirmed by measuring relative atomic displacements and piezoresponse force microscopy[24]. While the Dy-Dy pairs manifest as single columns in the image, which yields $d_{info} \approx 55$ pm, numerous O-O separations with $d_{O\text{-}O} \leq d_{info}$ are clearly resolved in the FE phase (Fig. 1f). With a minimum $d_{O\text{-}O} \approx 26$ pm, the maximum probability (MP) value in statistics is $d_{O\text{-}O\text{-}MP} \approx 50$ pm, and the O-column full width at half maximum (FWHM) is centered on $d_{FH} \approx 56$ pm (Fig. 1g and Supplementary Fig. S1). These results indicate that the Cs-TEM can achieve super-resolution imaging (SRI) of oxygen, and the $d_{info}$ can be attributed to the split O-O columns.

To find out origin of the SRI capability, we performed quantum mechanics-based multi-slice image simulations. At an optimal imaging condition of *Cs* = -10 um, *A1* = 1 nm, *A2* = 20 nm, *B2* = 20 nm, which corresponds to a Scherzer point resolution of $d_0 \approx 0.66\ \lambda^{3/4} Cs^{1/4} = 55$ pm, the specimen thickness is determined to vary in the range of $3.0 \leq t \leq 7.5$ nm (Fig. 1b and Supplementary Fig. S2a,b). Interestingly, the Dy-Dy split ($S_{\mathrm{Dy}}$ = 70 pm) can only be resolved at $t \leq 2$ nm thick. With increase of *t*, the Dy atom pairs evolve to an unresolvable state around $t = 4 \sim 10$ nm, after which the re-entrant Dy-Dy split jumps to ~130 pm and narrows down again with *t* (Fig. 1d,e). This suggests that the heavy Dy atoms may hierarchically focus the propagating electrons during the imaging process, which is supported by a V-shape extinction pattern along the depth direction (Fig. 1h and Supplementary Fig. S2c,d).

As reducing $S_{\mathrm{Dy}}$ from 70 to 0 pm, the extinction points shift upward and finally the V-shape pattern transforms to an O-shape chain (OSC) pattern. Irrelevant to B-O splitting along the viewing direction, our systematic study on a series of $ABO_3$ compounds such as $LiAlO_3$, $NaNbO_3$ and $ATiO_3$ (A = Ca, Sr, Ba) reveals that the V-shape extinction prevails not only in B/O columns (Fig. 1h), but also in heavy A-site atom columns, e.g., Z ≥ 38 (Sr) (Supplementary Fig. S3). This forms a sharp contrast to the conventional cognition, i.e., all atoms comply with the OSC-type extinction and their extinction distance ($\xi$) is proportional to $Z^{-1}$. As a matter of fact, the heavy-atom extinction is very sensitive to chemical coordination environment, which is exemplified by an abnormal extinction of Ba in $BaTiO_3$ and $BaBiO_3$ (Supplementary Fig. S3c,j)[23]. For light-atom oxygen, exhibiting either a small-angle divergence (SAD) or a parallel stripe (PS) shape (Fig. 1i), one general rule is that the atomic $d_{\mathrm{FH}}$ value in simulated images is always larger than $d_{\mathrm{YF}}$ and the experimental MP values (Fig. 1g).

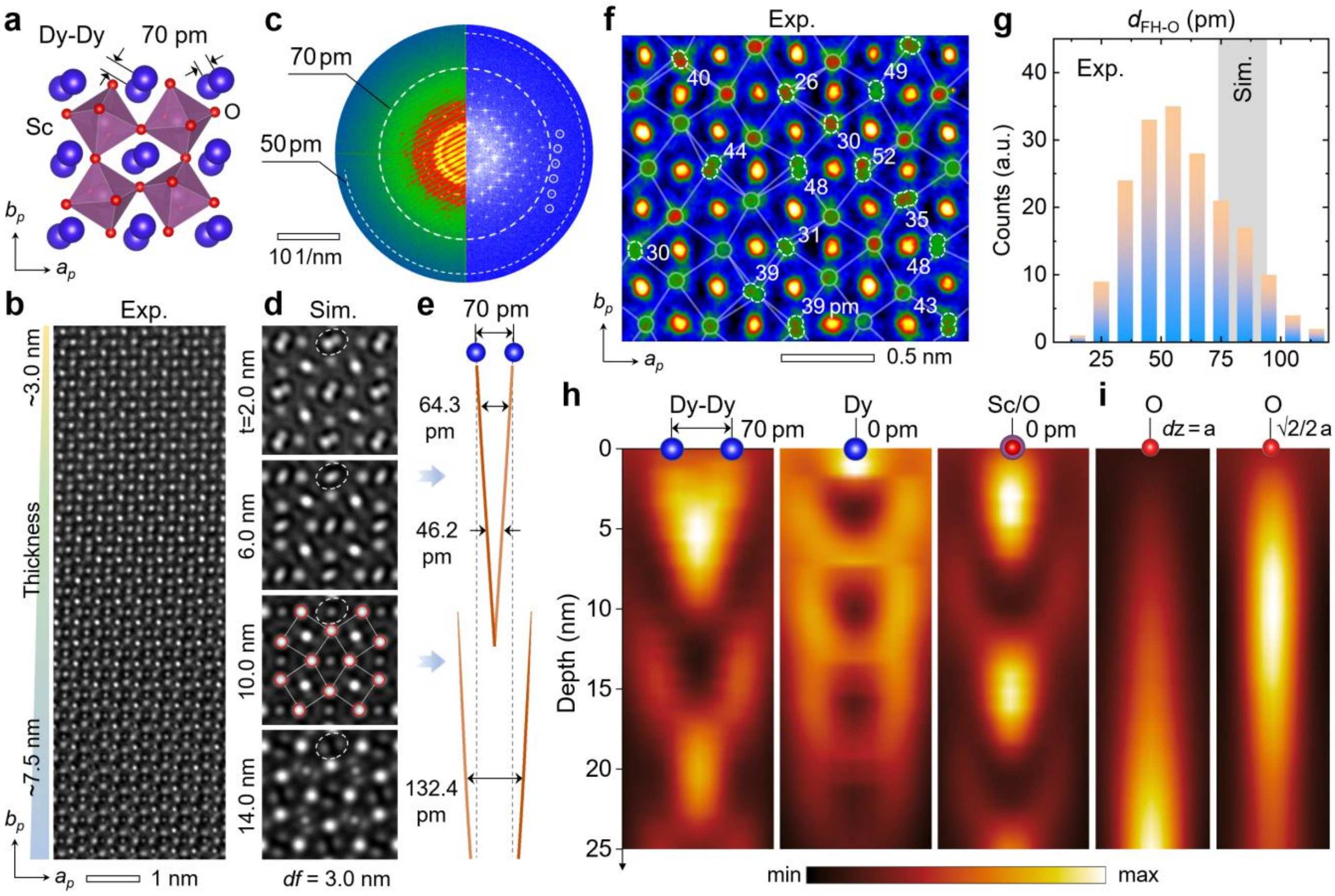

**Fig. 1| Element-dependent limited-resolution and super-resolution imaging in $DyScO_3$. a**, Crystal structure of $DyScO_3$ with Dy-Dy splitting ($S_{Dy}$ = 70 pm) and antiphase octahedral rotation ($a^-a^-c^+$ type). **b**, [001]-oriented atomic-resolution Cs-TEM image taken from a region with thickness gradient. **c**, Young's fringe diffractogram of our microscope measured at 300 kV (left) and a fast Fourier transform (FFT) image of (**b**, right) with highlight of 70 pm and 50 pm resolution rings. **d**,**e**, Simulated images with *Cs* = -10 um, defocus *df* = 3 nm, thickness *t* = 2.0 ~ 14.0 nm, and hierarchical focusing of Dy-Dy atoms manifested by varied Dy-Dy spacing, respectively. **f**, Experimental super-resolution oxygen imaging, with annotation of O-O splitting distance (in unit of pm). **g**, Statistical histogram of $d_{FH\text{-}O}$, with a peak around 56 pm, measured from (**b**) and that from image simulation (grey shadow). **h**,**i**, Depth-dependent interference patterns of channeling electrons for Dy-Dy ($S_{Dy}$ = 70, 0 pm), Sc-O ($S_{Sc\text{-}O}$ = 0 pm) and O ($dz = a$, $\sqrt{2}/2a$) atoms, respectively.

## Super-resolution hydrogen imaging

To validate the role of light atoms in SRI, we conducted comparative experiments on Ti and $TiH_x$, where the latter is obtained via high-temperature (200 °C) hydrogenation of Ti lamellae specimen in a $H_2$-gas environment. Our selected area electron diffraction experiments reveal that the H injection leads to a lattice expansion by ~5.4% (Fig. 2a,b and Supplementary Fig. S4a,b). As comparing their diffractograms, we find that the $d_{info}$ in Ti is 66.1 pm (15.13 1/nm), which steeply increases to $d_{info}$ =

55.7 pm (17.96 1/nm) in $TiH_x$ (Fig. 2c). With respect to Ti-atom arrangement in metal, one clearly observes hydrogen-atom contrast in structural images of $TiH_x$, either more in between adjacent Ti rows ($H_T$ site), or less in between adjacent Ti columns ($H_B$ site) (Fig. 2d,e). As a result, the Ti-Ti distance $d_S$ increases from 0.144 to 0.150 nm, and $d_L$ from 0.148 to 0.155 nm (s.d. ≈ 0.005 nm), respectively.

Alongside observation of H-H dimers with $d_{H\text{-}H} = 20.8 \sim 26.3$ pm, these results verify the critical role of hydrogen in realizing SRI. In statistics, the MP value of $d_{FH\text{-}H}$ appears at $d_{FH\text{-}H\text{-}MP} = 37.3$ pm (Fig. 2f,g). This forms a sharp contrast to a value of $d_{FH\text{-}H} \approx 112$ pm in the simulated image (Supplementary Fig. S3f). Specifically, we find that the effective H contrast, defined by a ratio of column peak intensity, i.e., $R_{H/X} = (I_H - I_{V_H})/(I_X - I_{V_H}) \times 100\%$ ($V_H$ stands for a H-free equivalent site), is as high as $R_{H/Ti} = 23.6 \pm 6.8\%$, which is ~ 5 times of the H contrast in $TiH_2$ obtained by electron ptychography[25]. This highlights a significant advantage of Cs-TEM in achieving high-contrast hydrogen imaging.

As a further proof of hydrogen intercalation, our electron energy loss spectroscopy (EELS) experiments reveal a blue shift of Ti-$L_{2,3}$ edges by 0.42 eV in $TiH_x$ against the Ti metal. This reveals that the hydrogenation increases the Ti valence state via charge transfer from Ti to H. Accordingly, we also observe a blue shift of bulk plasmon peak by 1.4 eV in $TiH_x$. Referring to previous EELS studies on $TiH_x$ ($0 \leq x \leq 2.0$)[26, 27], the relationship between plasmon peak shift and H content yields that x ≈ 0.6 in our sample (Supplementary Fig. S4c-e). Regarding the discrepancy of minimally resolved $d_{H\text{-}H}$ with $d_{info}$, it is worth emphasizing that only regular and periodic atomic arrangements can be manifested in a diffractogram. When such atomic separations adopt random orientation, finer structural motifs (e.g., $d_{H\text{-}H} \approx 20$ pm) cannot be reflected in the diffractogram.

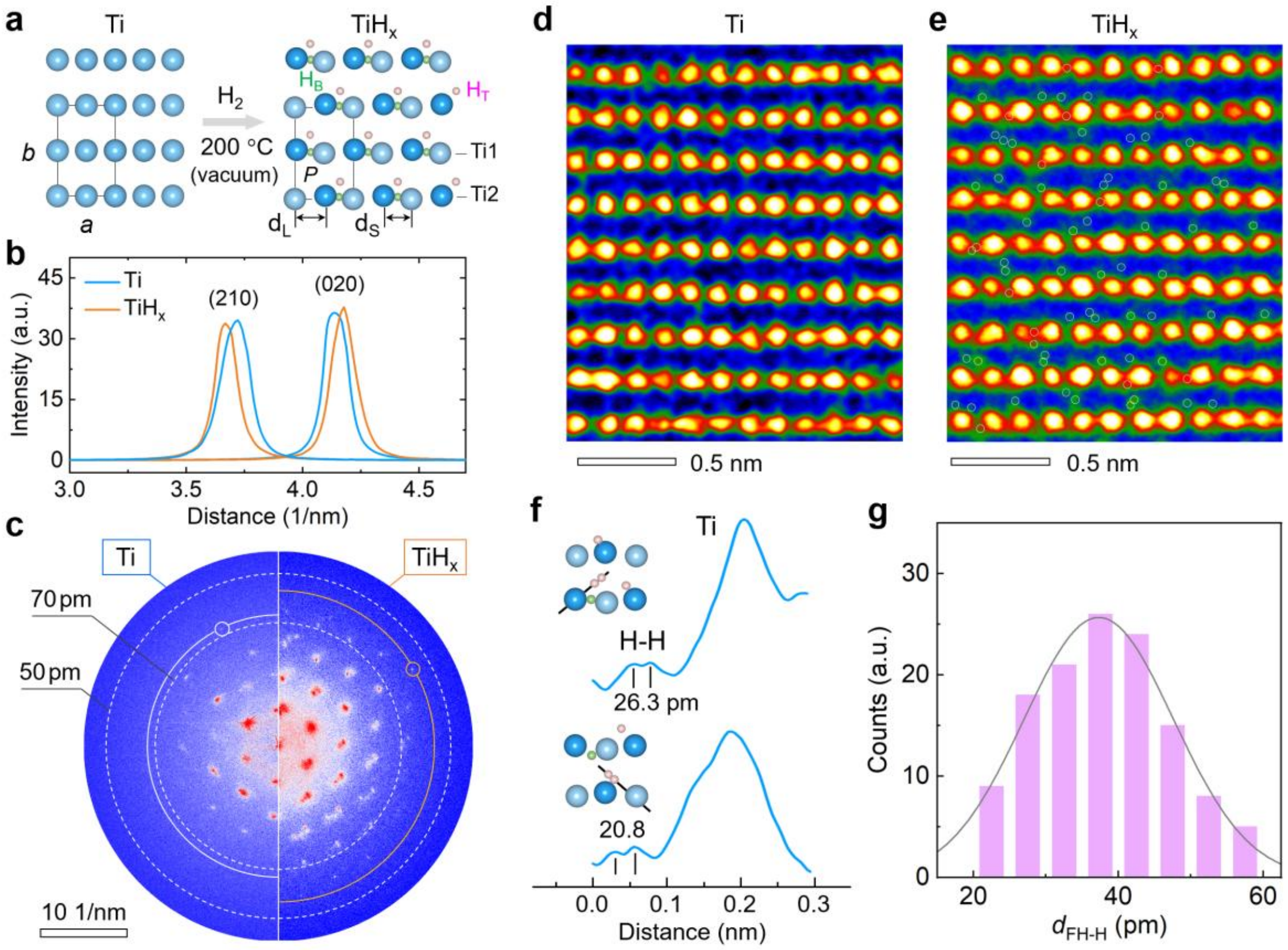

**Fig. 2| Super-resolution hydrogen imaging in $TiH_x$. a**, Illustration of Ti-to-$TiH_x$ transformation via hydrogenation at 200 °C in $H_2$ atmosphere. The bond length ($d_S$, $d_L$), Ti1 and Ti2 sites, $H_T$ (pink) and $H_B$ (green) sites are annotated. **b**, Intensity profiles of (210) and (020) diffraction for Ti and $TiH_x$ indexed based on the tetragonal structure in (**a**). **c**, Diffractograms of atomic-resolution TEM images of Ti (left) and $TiH_x$ (right) with highlight of their individual spatial cut-off frequency at 15.1 and 18.0 $nm^{-1}$, marked by solid circles, respectively. **d**,**e**, Local atomic-resolution Cs-TEM images of Ti and $TiH_x$, taken along [001] direction, with annotation of H atoms (white circles) in (**e**), respectively. **f**, Representative intensity profiles of H-H-Ti columns and corresponding sub-unit cell structures. **g**, Histogram of $d_{FH\text{-}H}$ for H-atom columns with the maximum probability value at $d_{FH\text{-}H\text{-}MP}$ = 37.3 pm.

**Particle nature dominated super-resolution imaging**

The above results demonstrate that the Cs-TEM can achieve SRI of light atoms. For heavy Dy, Sc/O and Ti species, their maximum probability $d_{FH\text{-}MP}$ values measured from the structural images can be nicely reproduced via multi-slice image simulations (Fig. 3a and Supplementary Fig. S2a-f, S4f). However, this is not the case for light O and H atoms. With decrease of Z, the experimental $d_{FH\text{-}MP}$ value decreases from 56.1 pm (O) to 37.3 pm (H). Opposite to this, their simulated value increases from ~85 pm (O) to ~104 pm (H) (Supplementary Fig. S3). Since the structural images are interpreted within the framework of wave-based imaging theory, which is founded on de Broglie's matter-wave hypothesis, the challenge of reproducing the experimental $d_{FH}$ values of light elements compels us to reconsider the particle nature of the imaging electrons.

It has been well established that the wave nature dominates as the wavelength ($\lambda$) is much larger than the characteristic size ($d$) of an object, i.e., $\lambda/d >> 1$, as exemplified by double-slit interference and low-energy electron diffraction (Fig. S5a). While for a Cs-TEM operated at 300 kV, the $\lambda/d << 1$ relation ($\lambda = 1.97$ pm, $d \approx 20 \sim 400$ pm) unveils that the particle nature of electrons cannot be ignored. By probing the propagation paths using a nearly parallel incident beam (radius ~4.3 nm), we find that the electrons travel ballistically along the atom column before reaching the extinction point (Fig. 3b and Supplementary Fig. S5b). In specific, the ballistic propagation is characteristic of very small $d_{FH}$ values. Taking z-direction O-O distance as a variable, as $dz$ reduces from $a$ to $a/4$ ($a = 0.4$ nm), one finds that the $d_{FH\text{-}O}$ is in the range of 29.6 to 46.6 pm. Since the propagation of electron particles is not subject to the modulation of contrast transfer function (CTF) of the objective lens, this provides a viable route to achieve SRI of oxygen (Fig. 3h).

Being analogous to our experimental observation, as reducing H-H distance from $dz = a$ to $a/8$, one finds that the ballistic electron propagation also yields much smaller $d_{FH\text{-}H}$ values. Taking $dz_{H\text{-}H} = a/8$ case as an example, we find that by keeping a nearly constant $d_{FH\text{-}H} \approx 35.6$ pm, which well agrees with the experimental $d_{FH\text{-}H\text{-}MP}$ value, the peak intensity $I$(z) can be expressed as

$$I(z) = A\cdot \sin^2(\pi\cdot z/\xi) + B\cdot \sin(2\pi\cdot z/\xi) + C. \quad (1)$$

Herein, our fitting yields $A \approx 0.024$, $B \approx 0.003$, $C = 0.035$, $\xi_H = 36$ nm (Supplementary Fig. S6). According to the established imaging theory[28], one can conclude that associated with direct transmission ($C$), the ballistic electron propagation described by the quadratic term governs the amplitude contrast of atom column[23], which increases with reduction of $dz$ (Supplementary Fig. S5c-e). For the linear interference term, ~13.5% of the quadratic one, it unavoidably widens the $d_{FH\text{-}H}$ value of H column and such a proportion increases with increase of $dz$ (Fig. 3c).

By comparing the $d_{FH}$ values from experiments, multi-slice simulation and particle propagation, the wave-particle ratio is estimated as $R_{WP} \approx 1/2$ and 0 for oxygen and hydrogen, respectively (Fig. 1f, 2g). This reveals that unlike the strong heavy atom-electron interaction, describable by wave-based imaging theory (Fig. 3f and Supplementary Fig. S1a-d), the weak Coulomb attraction of light atoms accentuates the particle nature of electrons. On this basis, we simulated a pair of oxygen columns with $dz_{O\text{-}O} = 2a$ and a mutual shift of $\Delta dz = a$ (Fig. 3d,g). As varying the $dx_{O\text{-}O}$ spacing, from 70 to 40 pm, the achievable O-O resolution in xy plane is ~45 pm, in good agreement with our experimental result on $DyScO_3$ (Fig. 1f). A similar resolution metric is also observed for a pair of H columns with $dz = 2\Delta dz = 3a/16 = 75$ pm, where the depth resolution is observed to reduce with $dx_{H\text{-}H}$ as well (Fig. 3e,g).

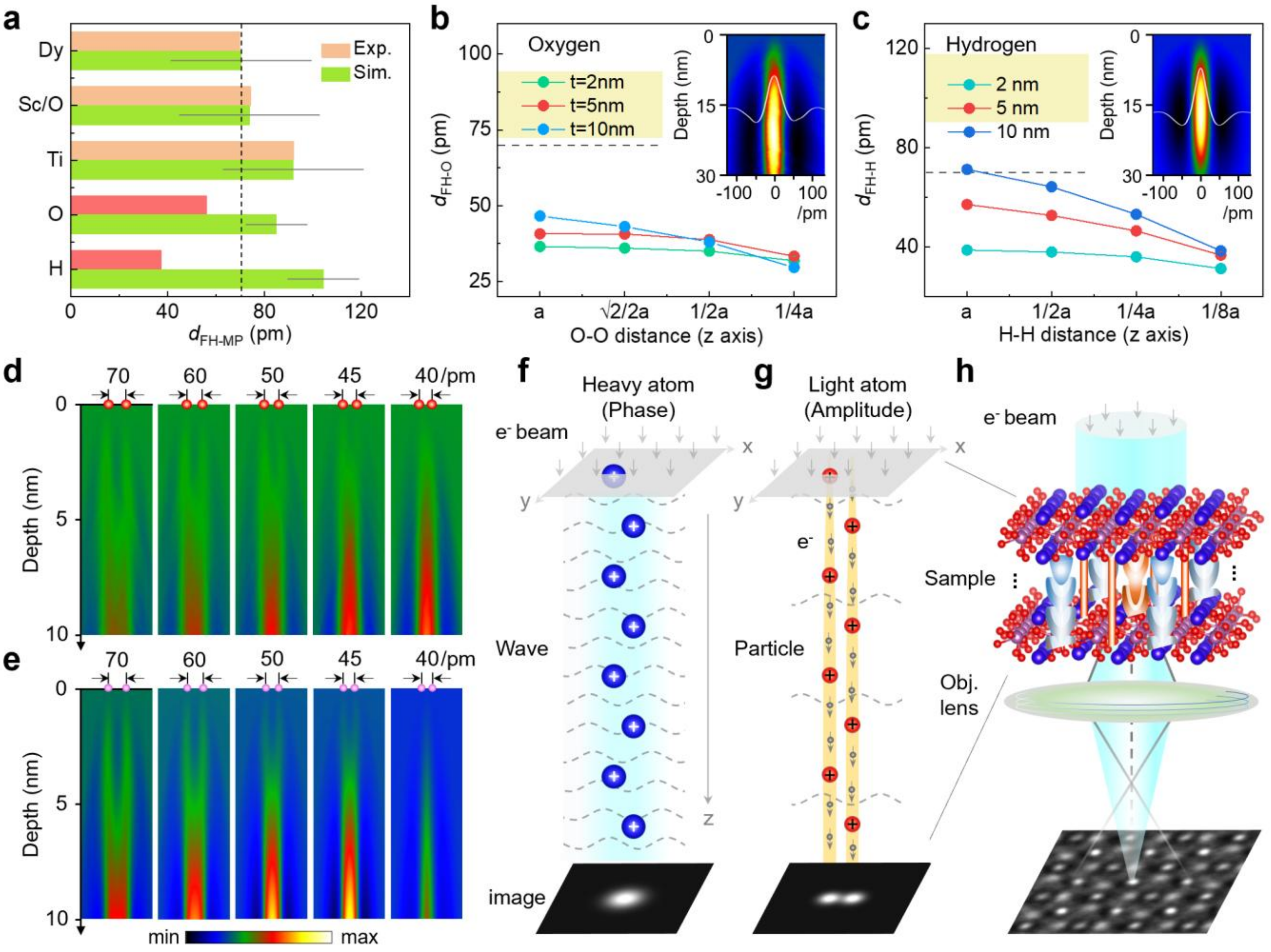

**Fig. 3| Ballistic propagation-mediated SRI of light elements.** **a**, Comparison of $d_{FH\text{-}MP}$ values measured from the experimental and multi-slice simulated images. The dashed line marks the $d_{YF}$. **b**,**c**, The $dz$-dependent $d_{FH\text{-}O}$ and $d_{FH\text{-}H}$ values at different thickness and representative propagation-path maps of electrons along oxygen (inset: $dz_{O\text{-}O} = a\sqrt{2}/2$, $a = 0.4$ nm) and hydrogen (inset: $dz_{H\text{-}H} = a/8$) columns, respectively. The yellow shadows denote the $d_{FH}$ ranges in wave-based simulations. The strong dependence of $d_{FH\text{-}H}$ on $dz$ is attributed to the increased proportion of linear interference term $\sin(2\pi z/\xi)$ against the ballistic propagation term $\sin^2(\pi z/\xi)$ in $I(z)$ (Supplementary Fig. S6). **d**,**e**, Electron propagation path maps for a pair of O-O ($dz = 2\Delta dz = 2a$) and H-H columns ($dz = 2\Delta dz = 3a/16$) with $dx$ reducing from 70 to 40 pm, respectively. Their z-direction period is $2a$ and $3a/16$,

respectively. **f-h**, Illustration of wave-based limited-resolution imaging of heavy atoms and particle-based SRI of light atoms in Cs-TEM, respectively.

## Super-resolution hydrogen imaging in perovskite oxides

As a further validation, we carried out comparative Cs-TEM experiments on complex perovskites $LaAlO_3$ and H-$LaAlO_3$, where the protons are introduced via ionic liquid gating (Supplementary Fig. S7a). For pure $LaAlO_3$, a thin-region diffractogram shows that the $d_{info}$ is 51.8 pm (19.31 1/nm). As it comes to H-$LaAlO_3$, the $d_{info}$ increases to 44.8 pm (21.85 1/nm), further confirming the critical role of H in improving the information resolution (Fig. 4a and insets). At the atomic scale, a Cs-TEM structural image of $LaAlO_3$ shows that alongside La-atom extinction, the oxygen columns show elliptical elongation (ellipticity ~1.25) due to O-O splitting ($a^-a^-a^-$) (Fig. 4b). While in H-$LaAlO_3$, the hydrogenation transforms the nonpolar $R\bar{3}c$ phase to a polar monoclinic one[29]. In specific, the polar shifts of O and La columns against Al/O columns are measured as $\delta x_{\text{O-Al/O}} \approx -10.6$ pm, $\delta y_{\text{O-Al/O}} \approx -2.5$ pm and $\delta x_{\text{La-Al/O}} \approx 6.9$ pm, $\delta y_{\text{La-Al/O}} \approx 1.4$ pm, which correspond to $P_x \approx 33.1$ and $P_y \approx 9.8$ μC cm$^{-2}$ (Fig. 4c and Supplementary Fig. S8a-c). The result is confirmed by our first-principles calculations, which shows that the H, with a content of ~ 2.0 H per unit cell, tends to locate within the Al-O (001) planes (Fig. 4g-DFT and Supplementary Fig. S7b,c). Our EELS experiments further reveal a redshift of La-$M_{4,5}$ edges by 0.20 eV and an increase of $M_5/M_4$ peak ratio from 0.88 to 0.95, indicating that the La valence state is slightly reduced[30] due to proton injection (Fig. 4h).

Specifically, we clearly resolve hydrogen columns at different interstitial sites (white circles) in the structural image, predominantly surrounding La and O columns along the vertical planes. Of particular interest is the observed aggregation of hydrogen into dimers, trimers and even polymers in the local unit cells (Fig. 4d). Quantitatively, the effective H contrast $R_{\text{H/X}}$ ranges from 19.3% to 25.3% (Supplementary Fig. S8d-f), which is ~ 5 times of that in metal hydride obtained by electron ptychography[13]. Meanwhile, our statistical analysis reveals that the minimum $d_{\text{H-H}}$, $d_{\text{H-H-MP}}$ and $d_{\text{FH-H-MP}}$ values are14 pm, 32.8 pm and 39.3 pm, respectively (Fig. 4d,f and Supplementary Fig. S9a-d). By combining wave- and particle-based simulations for $LaAlO_3$ and hydrogen, respectively, we reproduce the experimental image of H-$LaAlO_3$ (Fig. 4d,e). With a resolvable concentration ~0.5 H per unit cell, we found that the H-column intensity oscillates with $t$ as a result of heavy-atom extinction, where the interpretable specimen thickness can be up to ~25 nm. Due to affection of imaging artefacts, this thickness limit decreases to ~ 15 nm in $TiH_x$ (Supplementary Fig. S10,S11).

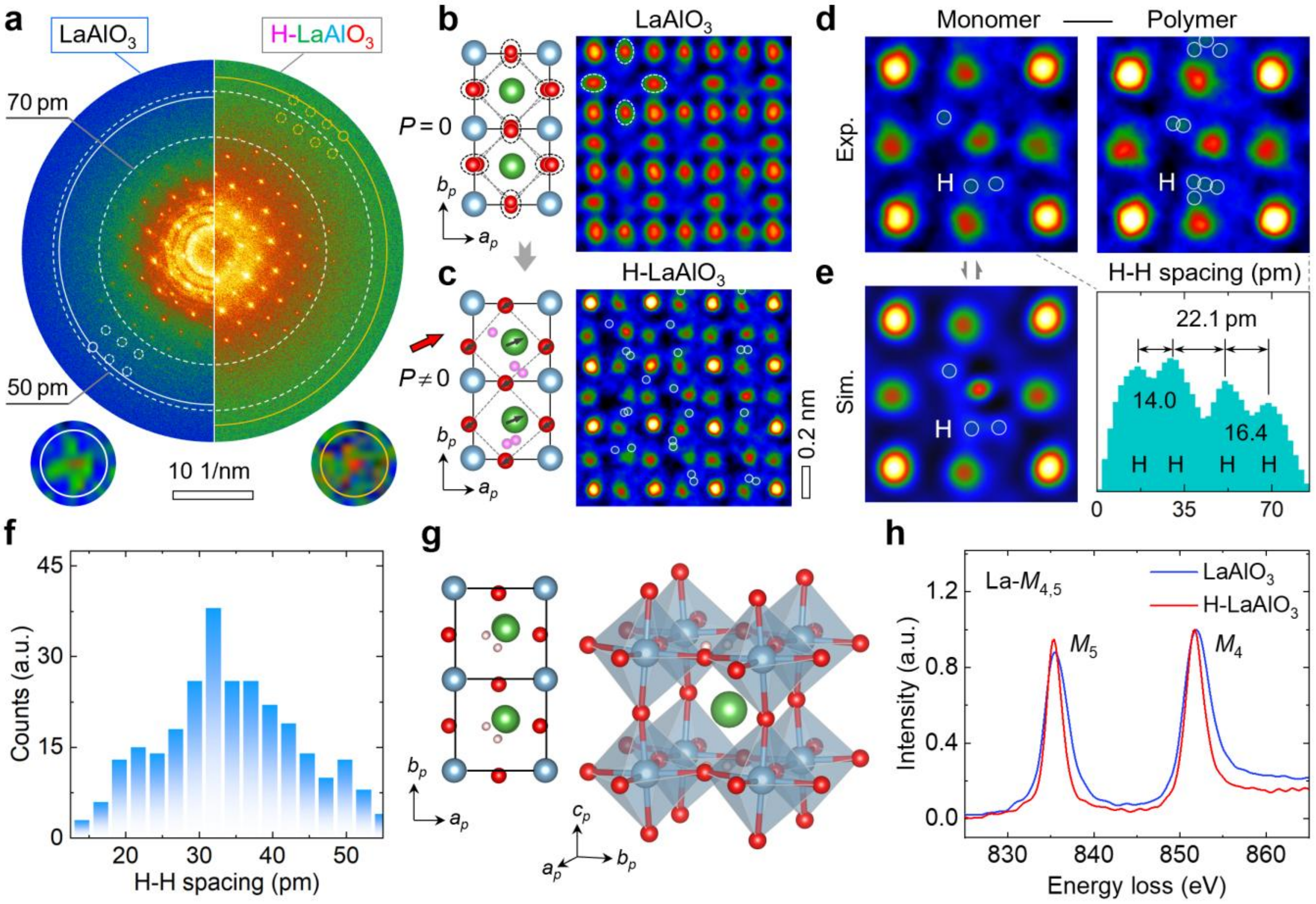

**Fig. 4| Super-resolution hydrogen imaging in H-LaAlO$_3$. a**, Diffractograms of LaAlO$_3$ (left) and H-LaAlO$_3$ (right) with spatial cut-off frequency (solid circles, insets) at 19.31 and 21.85 nm$^{-1}$, respectively. **b**,**c**, Schematic structures of nonpolar LaAlO$_3$ and polar H-LaAlO$_3$ (left) and their structural images (right) taken along [001] direction, respectively. The atom types are La-green, Al/O-blue, O-red, H-pink and white circles. **d**-**f**, La-centered experimental image, simulated image based on wave-particle duality (*Cs* = -5.5 nm, *t* = 4.9 nm, *df* = 2.3 nm), H-H intensity profile and statistical histogram of H-H spacing, respectively. With ~0.8 H per site, the H content in (**c**) is ~2.0 per unit cell. **g**, First-principles calculation of $H_{2.0}LaAlO_3$ with highlight of H location and octahedral shift-induced structural phase transition. **h**, Comparison of La-$M_{4,5}$ EEL spectra between LaAlO$_3$ and H-LaAlO$_3$.

## Hydrogen migration dynamics in complex oxide

Since the Cs-TEM is exceptional in temporal resolution, we took time-dependent snapshots to investigate the migration dynamics of hydrogen in H-LaAlO$_3$. As a function of beam irradiation time, we find a series of interesting behavior about local hydrogen migration. This includes a dimer-to-trimer transformation via adding a proton column, (12 $\rightarrow$ 24 s) at ($i$ = 5, $j$ = 3), transient hydrogen aggregation, (24 $\rightarrow$ 36 s) at ($i$ = 1, $j$ = 7), inter-site migration, rotational displacement and reorganization, (36 $\rightarrow$ 48 s) at ($i$ = 3, $j$ = 2), as well as hydrogen tunneling by up to ~1.54 nm of 4 unit cells (36 $\rightarrow$ 48 s) at ($i$ = 1 to 5, $j$ = 8) along the horizontal direction. Here, the hydrogen atoms closely adjacent to $La^{3+}$ and $O^{2-}$ ions can be roughly assigned as $H^-$ (green) and $H^+$ (pink), while the $H^0$ (white) species can act as charge-transfer intermediates to bridge their dynamic migration in the lattices (Fig. 5a-c).

More interestingly, our investigation reveals that the electron beam can drive collective H-occupancy transition via H tunneling. Statistical polar plots at $t_0$ = 0 and 12 s show that most H atoms are located at interstitials of lower-side La-$O_e$ bonds and La-Al/O bonds (quadrant, Q4), the remainder are close to La (Q2) and Al/O atoms (Q1), respectively (Fig. 5d). At $t_0$ = 24 s, the H occupancy shifts abruptly from between the lower-side La-$O_e$ bonds to the upper-side La-$O_e$ bonds and between the La-Al/O bonds in Q3 (Fig. 5e). This suggests that fractional incident electrons are captured by $H^+$-to-$H^0$ and $H^0$-to-$H^-$ conversion as propagating along the H columns. At $t_0$ = 48 and 60 s, the H occupancy recovers to its initial state (Fig. 5f). For the $t_0$ = 12 $\rightarrow$ 24 s transition, accompanied with negligible lattice change, the La and $O_a$ atoms abruptly displace further by $\Delta\delta_{La} \approx 15.6$ pm rightward and by $\Delta\delta_{Oa} \approx 17.2$ pm to the bottom left, respectively (Supplementary Fig. S12). This surges the unit-cell polarization to $P_x \approx 64.3$ and $P_y \approx 26.5$ μC $cm^{-2}$, about twice of its initial state (Fig. 5g). Such a dramatic change can be ascribed to transient storage of electrostatic energy[31] in insulating H-$LaAlO_3$.

As analyzing the H-X bond change, one finds that stimulated by increased H-H bond at $t_0$ = 12 s, the La-H and O-H bonds jointly increase from ~76 pm at $t_0$ = 0 s to ~ 95 pm at $t_0$ = 24 s (Fig. 5h,i). Following this, the La-H bond length step-wisely drops to 79.9 pm at $t_0$ = 60 s, while the H-H bond length fluctuates to a large value again. Accordingly, we noticed that the contrast of several La-atom columns is enhanced (upper dashed circle and square). Our image simulations indicate that such a change can be attributed to stripping of $LaH_x$ species from the specimen (Supplementary Fig. S13). The dynamic process can be visualized from the time-dependent structural images, e.g., (24 $\rightarrow$ 36 s) at ($i$ = 4, $j$ = 7), (36 $\rightarrow$ 48 s) at ($i$ = 1, $j$ = 7) and ($i$ = 3, $j$ = 3). In comparison, the oxygen atoms are reluctant to be stripped off during this process.

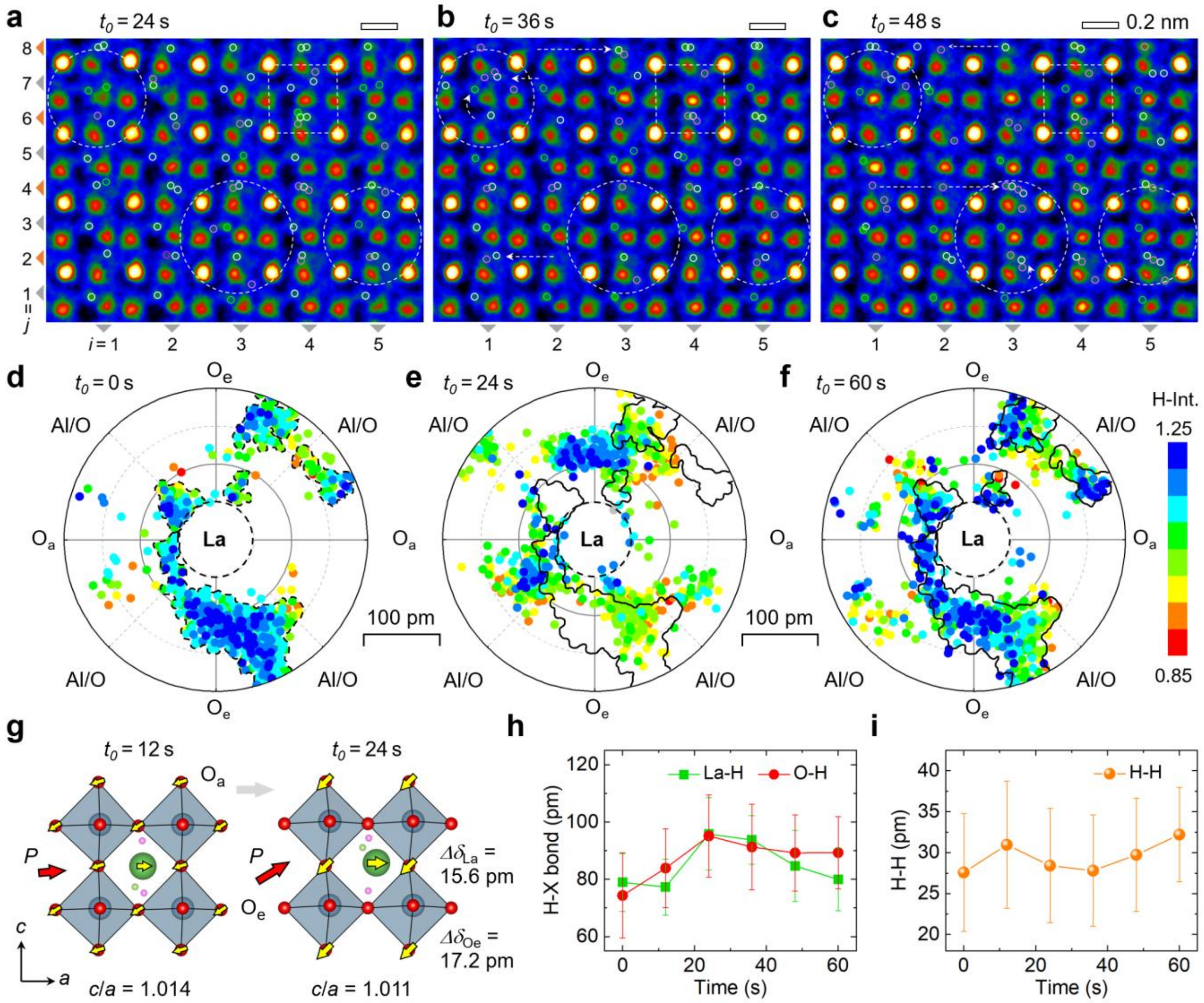

**Fig. 5| Hydrogen migration dynamics in H-LaAlO₃. a-c**, Time-resolved structural images obtained by Cs-TEM at $t_0$ = 24, 36 and 48 s along [001] direction under a dose rate of $8.8 \times 10^4$ eÅ$^{-2}$s$^{-1}$, respectively. The small pink, green and white circles mark the probable $H^+$, $H^-$ and $H^0$ species and their dynamic activities involving inter-site migration, rotational displacement and tunneling (arrows), within specific unit cells (dashed circles and squares). **d-f**, Time-resolved statistical evolution of H-atom distributions within La-atom-centered pseudocubic unit cells at $t_0$ = 0, 24 and 60 s. **g**, Illustration of hydrogen occupancy transition induced by electron capture via the $H^+$/$H^0$-to-$H^0$/$H^-$ conversion from a low-energy state ($t_0$ = 0, 12 s) to a high-energy state ($t_0$ = 24 s), as well as collective hydrogen tunneling induced additional atomic displacements (thick yellow arrows) and polarization surge (red arrows). **h**,**i**, Time dependent length variation of La-H, O-H, H-H bonds as a function of time, respectively.

## Discussion and outlook

For a long time, the $d_{\mathrm{YF}}$ and $d_{info}$ have frequently been conflated. Since the adoption of aberration correctors, nonlinear electron-matter interactions involving dynamical scattering of heavy atoms[32] have been invoked to reconcile their difference. Herein, this work clarifies that the $d_{YF}$ actually evaluates the imaging ability of electron wave and the resolution metric, as a matter of fact, is elemental dependent in the measurement. For specimens containing heavy atoms only, the microscope resolution can be well described by classical wave-based imaging theory. While as light atoms are involved, the electron particle-dominated imaging bypasses the modulation of CTF and gives rise to the $d_{info}$. Since non-periodic atomic separations are not encoded in diffractograms, their finer structural details can only be retrieved from real-space structural images, wherein the minimum interatomic distance defines the resolution of SRI.

**Table 1.** Comparison of H-atom imaging effect between phase-contrast STEM techniques and Cs-TEM in $TiH_x$.

| Category | ABF | DPC | Ptychography | Cs-TEM (This work) |
|---|---|---|---|---|
| Ease of identification | ★★⯪☆☆ | ★☆☆☆☆ | ★★⯪☆☆ | ★★★★★ |
| SNR | 6.30 | 2.48 | — | 147.45 |
| H-atom contrast | $0.21 \pm 0.14$ | $0.22 \pm 0.16$ | $0.03 \pm 0.01$ | $0.24 \pm 0.06$ |
| Location accuracy | ~6.0 pm | ~8.0 pm | ~4.0 pm | ~3.0 pm |
| $d_{FH\text{-}H}$/ pm | $36.8 \pm 10.6$ | $40.9 \pm 14.3$ | $52.6 \pm 7.9$ | $37.5 \pm 6.7$ |
| Temporal resolution | ~ 8 s | ~ 5 s | 2 ~ 60 s | 0.001 ~ 10 s |
| Thick. endurance | ~10 nm [10,11] | ≤ ~ 5 nm [33] | ~ 20 nm [25] | ~15 nm |
| Typical image of $TiH_x$ ($x \approx 0.6$) | 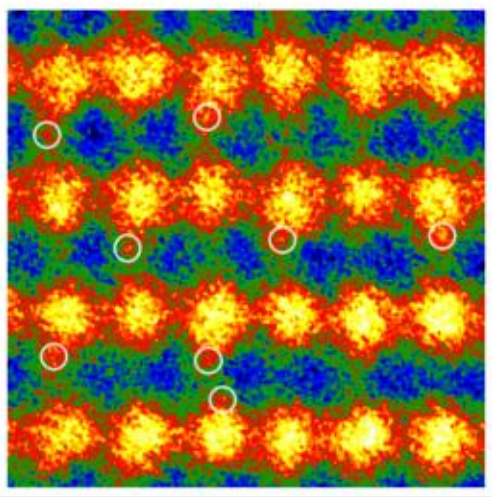 | 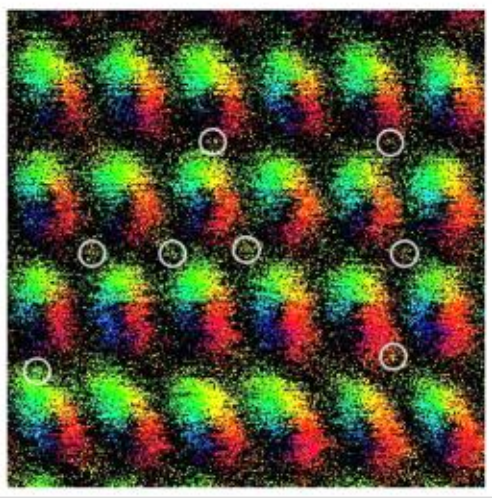 | 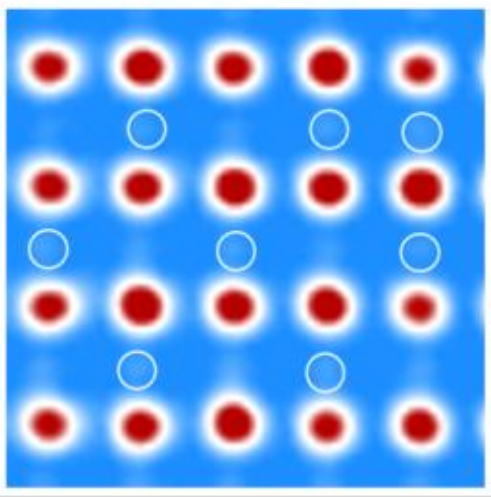 | 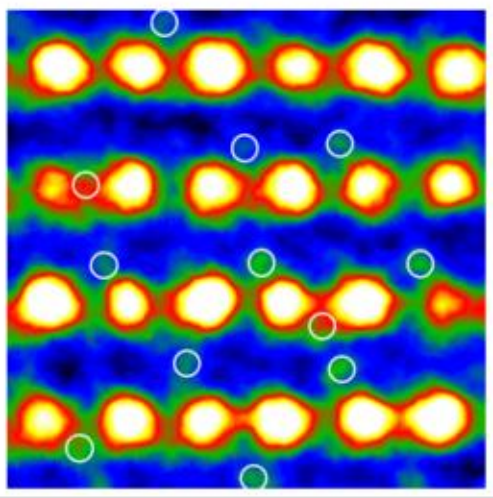 |

To demonstrate the superiority of Cs-TEM in hydrogen imaging, we acquired ABF-, DPC- and 4D-STEM datasets of $TiH_x$ for comparison. Our quantitative image analysis reveals that the scanning-induced noise greatly degrades the signal-to-noise ratio (SNR) of ABF and DPC images, which is two orders of magnitude lower than that of Cs-TEM. As a result, this not only hinders the identification of hydrogen columns, but also reduces their measurement accuracy (Supplementary Fig. S14). Notably, the H-column contrast and $d_{FH\text{-}H}$ values from ABF and DPC images closely match that of the Cs-TEM image, implying a common mechanism in hydrogen imaging. In sharp contrast, our ptychography study yields $R_{H/Ti}$ ≈ 3% and $d_{FH-H}$ ≈ 52.6 pm, in good agreement with $R_{H/Ti}$ ≤ 5% and $d_{FH-H}$ ≈ 63.6 pm reported in another ptychography study of $TiH_2$[25]. Together with thickness endurance[10, 11, 33] and temporal resolution (Table 1 and Supplementary Text 1), these quantitative metrics underscore that Cs-TEM offers unparalleled advantages for probing atomic-scale hydrogen dynamics. The conclusion is further reinforced by our comparison of ABF-, DPC-STEM and Cs-TEM data for the complex H-$LaAlO_3$ system (Supplementary Fig. S15).

## Conclusions

By addressing the long-standing resolution dilemma, our systematic study demonstrates that the Cs-TEM can achieve SRI for light elements. Essentially, this arises from the ballistic propagation of particle-like electrons boosted by weak Coulomb attraction of light atoms, which bypasses the objective-lens CTF constraints and surpasses the resolution limit imposed by wave nature of electrons. Thanks to the robust image contrast, immune to amorphous-layer noise and ordinary dose rates[34, 35], the Cs-TEM yields an ultrahigh hydrogen contrast of ~24% and the resolution for H-H separation is as low as 14 pm. In sharp contrast, the strong interactions of heavy atoms with electrons, encompassing dynamical scattering, diffraction and interference, can be well described by the wave nature of imaging electrons. With decrease of Z, e.g., from Z = 8 (O) to 1 (H), the wave/particle ratio is found to decrease as well. The transformative advance in harnessing electron-matter interactions not only surmounts the long-standing technical barriers for dynamic hydrogen imaging, but also offers new insights into phase-contrast STEM imaging.

On this basis, our in-situ Cs-TEM study of H-$LaAlO_3$ reveals, for the first time, polymeric hydrogen aggregation, dynamic hydrogen migration and collective hydrogen tunneling-induced large heavy-atom displacements and polarization surge. The direct observation of hydrogen migration dynamics and guest-host interactions establishes a paradigm for constructing precise structure-property relationships. In contrast to phase-contrast STEM techniques, limited by multiple constraints, the SRI in Cs-TEM offers superior temporal resolution, facile image interpretation as aided by an AI image simulator (see Methods) and low data volume, paving the way for its broad deployment. When coupled with spectroscopy, 3D tomography and external-field stimuli (thermal, electric, mechanical and optical), we firmly believe that the SRI holds great potential to uncover a suite of emerging phenomena and mechanisms across diverse functional and energy materials. This includes, but not limited to, metal hydrogen embrittlement, catalytic spillover, superconductivity and proton transport in fuel cells[14, 36, 37, 38].

# Methods

## Materials fabrication

The epitaxial thin films of $NdNiO_3$/$LaAlO_3$, with thickness of the $NdNiO_3$ layer at ~20 nm, were grown on single crystal substrates by using a pulsed laser deposition system. The growth conditions were optimized at 600 °C under an oxygen atmosphere of 10 Pa, respectively. The laser energy (KrF, $\lambda$ = 248 nm) was maintained at 2.0 J $cm^{-2}$, with a repetition rate of 3Hz. After the thin-film growth, the samples were cooled down to room temperature at a rate of 10°C $min^{-1}$ in an environment of 300 mbar to prevent overoxidation. The thin-film thickness was determined by X-ray reflectometry measurements and the results along with other physical properties were reported elsewhere [39]. The $TiH_x$ sample for TEM investigation was fabricated by hydrogenating a Ti lamellae specimen, which was prepared by a focused ion beam (FIB) system, at high temperature in a $H_2$-gas environment.

## Hydrogenation to the target materials

The Ti lamellae specimen prepared by FIB was thermally annealed at 200 °C in a 5 vol% $H_2$ Ar mixed atmosphere (0.15 MPa), where the annealing chamber was purged several times to remove the gas contamination. Upon hydrogen reduction, the $TiH_x$ lamellae specimen was cooled down to room temperature in the $H_2$-Ar atmosphere for TEM study. For directional hydrogen injection to the oxides, the thin-film edges were painted with a silver glue as the bottom electrode, the platinum wires fixed on the conductive electrode by silver glue was used as the gate electrode. A small drop of ionic liquid (DEME-TFSI) was added to cover both the film surface and the screwed platinum wire. The voltage between the bottom and gate electrodes was ramped from 0 V to 3 V in several minutes. It takes several tens of minutes to complete hydrogen injection to the $NdNiO_3$/$LaAlO_3$ [39] systems.

## Cross-sectional lamella specimen preparation

The lamellae specimens were prepared by using focused ion beam (FIB, Carl Zeiss CrossBeam 350) system. After lifting out and welding the specimens to the Omni-probe grids, the specimens were milled down by using Ga ions. In this process, the voltage is reduced from 30 kV to 5 kV and the ionic beam current is reduced from 2.8 nA to 90 pA. Following this, NanoMill 1040 (Fischione) operated at 500 V was used to further thin down the specimen and remove the surface contamination at liquid-nitrogen temperature.

**Image-corrected TEM experiments**

The atomic-resolution Cs-corrected TEM experiments were performed on a JEOL JEM-ARM300F2 microscope, which is equipped with a cold field emission gun (operated at 300 kV), double Cs correctors for the objective and condenser lens, a OneView camera, a GIF system for electron energy loss spectroscopy (EELS) and a couple of ultra-large area (316 $mm^2$) SDD system for energy dispersive spectroscopy of X-ray. The morphology and electron diffraction experiments were performed on a JEOL JEM-F200 microscope. Before acquiring atomic-resolution Cs-TEM images, the microscope was tuned to the best imaging conditions with $Cs \approx -10$ µm, $A1 < 2$ nm, $A2$ and $B2 < 30$ nm, etc., the acquisition time for each image frame was set as 0.6 s. For atomic position measurement, two-dimensional Gaussian fitting to peak values of the columns were performed on the basis of maximum likelihood estimation method. The electron propagation path simulation was carried out under STEM mode using Dr. Probe. Being analogous to TEM imaging, the nearly-parallel incident beam has a semi-convergent angle of 0.28 mrad and the beam with a diameter of ~ 4.3 nm is used to collect the propagation of electrons along the atom column.

**Wave-based and particle-based image simulations**

We developed an AI-based iterative image-matching algorithm in Python to rapidly determine the experimental imaging conditions. Alongside structural modelling using Vesta and CrystalKit, this algorithm produced simulated image sets by scanning the imaging parameters, where the multi-slice simulations were carried out in Dr. Probe and the modulation transfer function of OneView camera was incorporated. Essentially, the atomic quantification process involves three main steps. Firstly, to enable quantitative comparison, the experimental and simulated images were normalized to a unified grayscale intensity range. This eliminates biases from absolute intensity differences and preventing spurious luminance deviations in subsequent similarity evaluation. Secondly, we introduced a dynamic weighting scheme to balance the accuracy and computational efficiency, where the contribution of luminance $l(x_i, y_i)$, contrast $c(x_i, y_i)$ and structure similarity $s(x_i, y_i)$ were adapted to the statistical analysis. This adaptive coupling ensures the $SSI_M$ metric[40] reliably reflects physically relevant parameter variations, with the optimal imaging conditions determined by the maximum $SSI_M$ value. Thirdly, the best-fit atom-column peak intensities were then correlated with the mean-renormalized experimental counterparts to derive a global intensity correction factor.

By combining structural model refinement, including Debye-Waller factors, the hydrogen-free experimental image can be nicely reproduced by the wave-based image simulations. Benefitting from this, the image simulation efficiency is enhanced by two orders of magnitude, from about a month to a few hours. For the multi-slice simulated image $I_a(x,y)$, we note that it has a mean value approximately unity, which is independent of any imaging parameters. Similarly, the single-pass electron propagation simulation image $I_b(x,y)$, containing hydrogen columns only, also has a fixed mean value. After normalizing $I_b(x,y)$ by its mean value and subtracting the unity, the net H intensity was superimposed to the wave-based simulation image. By tuning the H-atom amount in structural model and corresponding intensity, we reproduced the experimental super-resolution image via the third step and obtained the H concentration in each simulated unit cell. The result ~ 2.5 H per unit cell well agreed with the reported one, ~2.2 H per unit cell in $SrCoO_{2.5}$, determined by TOF-SIMS[41].

**Probe-corrected STEM and EELS experiments**

The ABF, DPC, electron ptychography and EELS experiments were performed on the JEOL JEM-ARM300F2 microscope, operated at 300 kV. With a convergence semi-angle of 24 mrad, the ABF and DPC signal collection was around 10 ~20 mrad, where their acquisition time is 8.39 s (dwell time 2 μs/pixel, 2048 × 2048 pixels) and 5.24 s (5 μs/pixel, 1024 × 1024 pixels) for each frame, respectively. By utilizing the STEMx modular, the ptychography dataset was acquired using a Gatan-K3 camera over a 128 × 128 scanning grid, with a beam current of ~7 pA, a step size of 0.04 nm and an acquisition time of 49 s (dwell time 3 ms/pixel). Under an over-focus condition of *df* = 10 nm, the central-beam overlap was estimated as ~91.7%. The diffraction patterns were recorded by a 256 × 256 pixel-array detector. We reconstructed the dataset by using the PtyRAD algorithm with a multi-slice forward model[42]. In total, the object was cut into 27 slices and each one has a thickness of 1.5 nm. With a dispersion of 0.18 eV per channel, the EELS signal collection time was set as 0.3 ms per pixel, and the total number of scanning cycles was set as 30 to improve the signal-to-noise ratio.

**Time-of-flight secondary ion mass spectroscopy**

The TOF-SIMS data was obtained by IONTOF M6 instrument, where the mass resolution is ~ 200 atomic mass units. The Cs-ion sputtering beam (1 keV) was employed to produce a crater with a size about 500 μm × 500 μm and the negative secondary ions were collected from a central analysis area (100 μm × 100 μm) to minimize disturbances emanating from the edge of the crater. The AlO species was used as a reference to evaluate the concentration change of hydrogen between the pristine $LaAlO_3$ and H-$LaAlO_3$ crystals.

**Polarization estimation**

The spontaneous polarization in H-$LaAlO_3$ was calculated by using the formula: $P_s = \frac{1}{V}\sum_i \delta_i Z_i$, where the $V$, $\delta_i$ and $Z_i$ are the unit cell volume, displacement of the $i$-th atom and the Born effective charge of the $i$-th atom, respectively. The effective charges used in the calculation are $Z_{La} = 4.45$, $Z_{Al} = 2.91$, $Z_{Oa} = Z_{Oe} = -2.44$ [43], with Oa and Oe denote the apical and equatorial O site, respectively. The off-center displacement $\delta_i$ is relative to the A-site atom and the contribution of hydrogen is not considered in our polarization estimation.

**First principles calculations**

The crystal structures were optimized by density functional theory (DFT) calculations using the VASP code [44, 45], which employs the projector augmented-wave (PAW) method[46, 47]. The exchange-correlation energy was treated within the $r^2$SCAN meta-generalized gradient approximation (meta-GGA) functional[48]. The PAW-PBE potentials were used for La, Al, O, and H. A plane-wave basis set with a kinetic energy cutoff of 520 eV was employed, and the convergence criterion for the total energy was set to $10^{-6}$ eV. The Monkhorst-Pack sampling was adopted for Brillouin-zone integration[49], with a k-point spacing of $2\pi \times 0.033$ Å$^{-1}$ for structural relaxation and $2\pi \times 0.025$ Å$^{-1}$ for subsequent self-consistent calculations. During structural relaxation, all atomic coordinates and the cell volume were optimized under a fixed cell shape constraint, with a force convergence criterion of 0.01 eV Å$^{-1}$. In supercells containing eight $LaAlO_3$ formula units, we found that the experimental polar structure of H-$LaAlO_3$ can be well reproduced as the H content is up to 2H per unit cell, where the H-H bond length is ~ 0.75 Å. On this basis, the calculated band structure showed that the band gap was reduced from 4.4 eV in $LaAlO_3$, smaller than the experimental value (~5.6 eV)[50], to 3.06 eV in H-$LaAlO_3$.

## Data availability

All the experimental and simulation data needed to evaluate the conclusions in the paper are present in the paper or Supplementary information.

## Acknowledgements

This work was supported by the National High-Level Youth Talents Program (Grant Nos. 0040/X2450224, 0040/K2923004), the Xiamen Natural Science Foundation Program (Grant No. 3502Z202573024), the National Natural Science Foundation of China program (Grant No. 52573256), the Xiamen Major Science and Technology Program (Grand No. 3502Z20261015), and the State Key Laboratory of Vaccines for Infectious Diseases of Xiang An Biomedicine Laboratory (Grant No. 2024XAKJ0100006).

## Author contributions

X.-K.W. conceived the research idea and coordinated the study. Z.Liang, F.L., L.C. carried out the TEM experiments, quantitative data analysis, image simulations and algorithm development. Z.Lv, and X.L conducted the ptychography experiments and simulation of interference map. S.C. and Y.S. performed first-principles calculations. N.L., Y.H. and Q.Z. provided the H-$LaAlO_3$ samples and prepared the FIB specimens. Y.W. and P.L. performed the hydrogen injection into $LaAlO_3$ via ionic liquid gating and TOF-SIMS experiments. H.B.T. provided the pure Ti sample, Z.Liang and P.L. prepared the FIB specimens of $TiH_x$ samples. C.Z. and R.E. D.-B. provide technical support and helpful discussion. X.-K.W. wrote the manuscript with input and contribution from all co-authors.

## Competing interests

The authors declare that they have no competing interests.

## Additional information

Supplementary information.

Supplementary Materials

# Ballistic propagation-mediated super-resolution light-element imaging: in-situ tracking dynamic hydrogen migration

Zhiyao Liang,[1,7] Feng Liu,[1,7] Linyuan Chen,[1,7] Zhiwei Lv,[1] Shiya Chen,[2] Xinyu Liao,[1] Yuzhou He,[3] Yi Wang,[1] Pengpeng Liu,[1] Hua Bing Tao,[1] Nianpeng Lu,[3] Changlin Zheng,[4] Yang Sun,[2] Qinghua Zhang,[3] Rafal E. Dunin-Borkowski,[5] Xian-Kui Wei[1,5,6]*

[1]College of Chemistry and Chemical Engineering, Xiamen University, 361005, Xiamen, China

[2]Department of Physics, Xiamen University, Xiamen, 361005, China

[3]Beijing National Laboratory for Condensed Matter Physics, Institute of Physics, Chinese Academy of Science, Beijing 100190, China

[4]State Key Laboratory of Surface Physics and Department of Physics, Fudan University, Shanghai, 200433, China

[5]Ernst Ruska-Centre for Microscopy and Spectroscopy with Electrons, Forschungszentrum Jülich, Jülich, 52428, Germany

[6]Innovation Laboratory for Sciences and Technologies of Energy Materials of Fujian Province (IKKEM), Xiamen 361005, China

[7]These authors contribute equally: Zhiyao Liang, Feng Liu, Linyuan Chen

Corresponding author Email: xkwei@xmu.edu.cn

# Supplementary Information

## This PDF file includes:

Supplementary Text 1
Supplementary Figures 1-15
Supplementary References 1-12

# Supplementary Text

## 1. Comparison of STEM-vs-TEM imaging mechanism

For phase-contrast STEM imaging techniques, their distinction in signal collection leads to different dependence of atomic contrast on Z and specimen thickness. As far as metal hydrides are concerned, the heavy- and light-atom contrast in ABF-STEM is reported to originate from incoherent thermal diffuse scattering and electron channeling effect, respectively[1]. With a $Z^{1/3}$ dependence of atomic contrast, the H-atom imaging thickness limit is ~ 10 nm[2, 3]. For DPC-STEM, it directly images the projected electric field of atom columns, which follows a $Z^1$ dependence, and the interpretable specimen thickness is ≤ 5nm[4]. With removal of dynamical scattering effects, the interpretable H-atom contrast in ptychography is up to ~ 20 nm, where the atom contrast follows a $Z^1$ relation as well[5, 6].

Under the negative Cs imaging mode, the Cs-TEM delivers an especially high and robust image contrast[7]. This is achieved by superposing the phase contrast with the amplitude contrast, which arises from the diffraction channeling effect of electrons[8, 9]. In our study, we reveal that the weak Coulomb attraction between light atoms and electrons promotes the ballistic propagation of particle-like electrons. This effect coexists with the strong Coulomb interaction between heavy atoms and electrons, which can be well described by the wave-based imaging process. As a result, the enhanced amplitude contrast gives rise to ultra-narrow $d_{FH}$ values and the SRI of light atoms. Specifically, on the one hand, the heavy atoms may extinct within the effective detection thickness (up to ~15 nm), i.e., the atomic contrast roughly follows a $Z^{-1}$ relation, thus the light-atom contrast is highlighted accordingly. On the other hand, the negative Cs imaging mode is immune to noise signal from surface amorphous layers[7] and relatively low electron dose rate[10]. Therefore, the Cs-TEM provides unparalleled and all-round technical advantages in conducting studies about hydrogen dynamics.

## Supplementary Figures

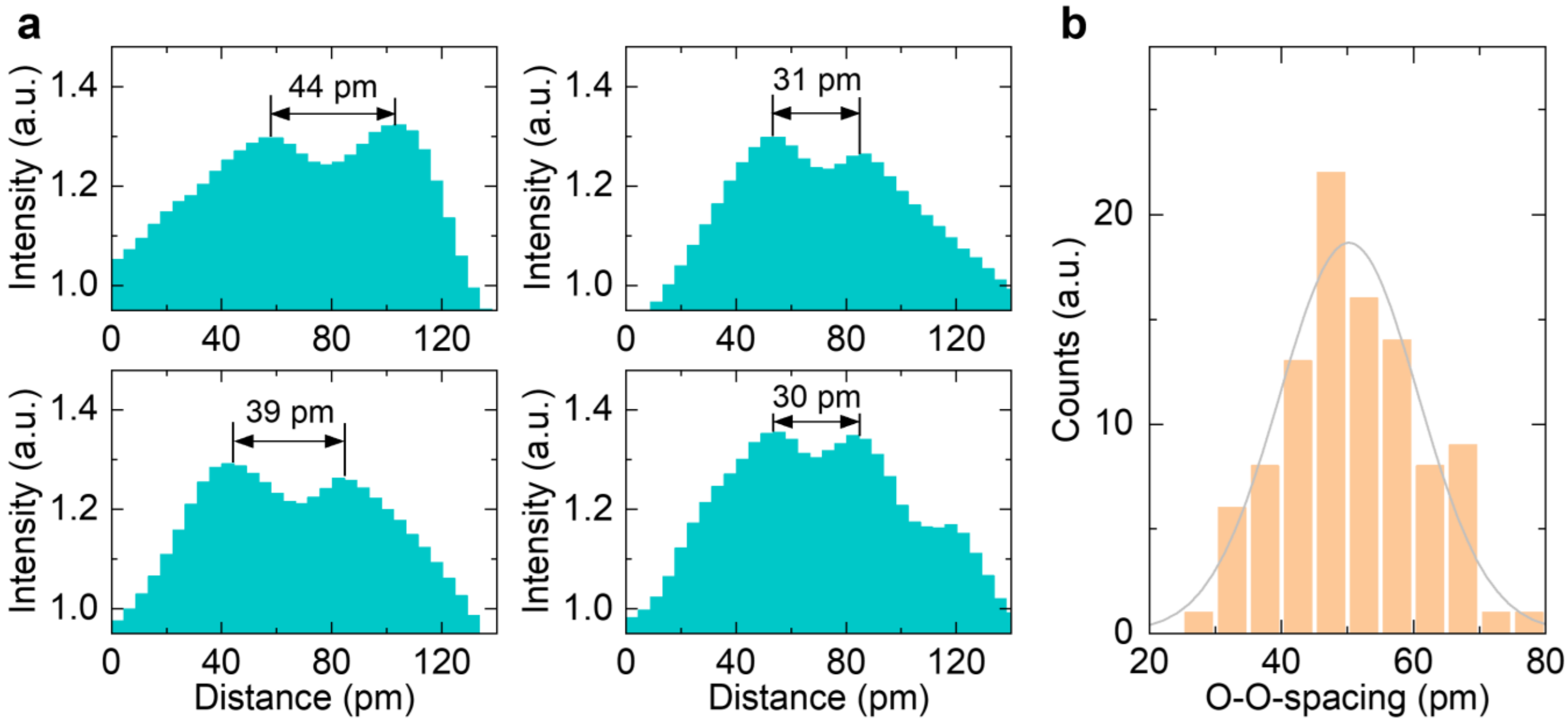

**Supplementary Fig. 1. Experimental measurement of O-O splitting from the FE phase of $DyScO_3$.** **a**, Representative intensity profiles of split O-O columns measured from the experimental image of Fig. 1b. **b**, Statistical histogram of O-O spacing with a Gaussian fit. With $d_{FH\text{-}O}$ = 42.6 ± 12.3 pm, the MP value appears at $d_{O\text{-}O\text{-}MP}$ = 50.2 pm.

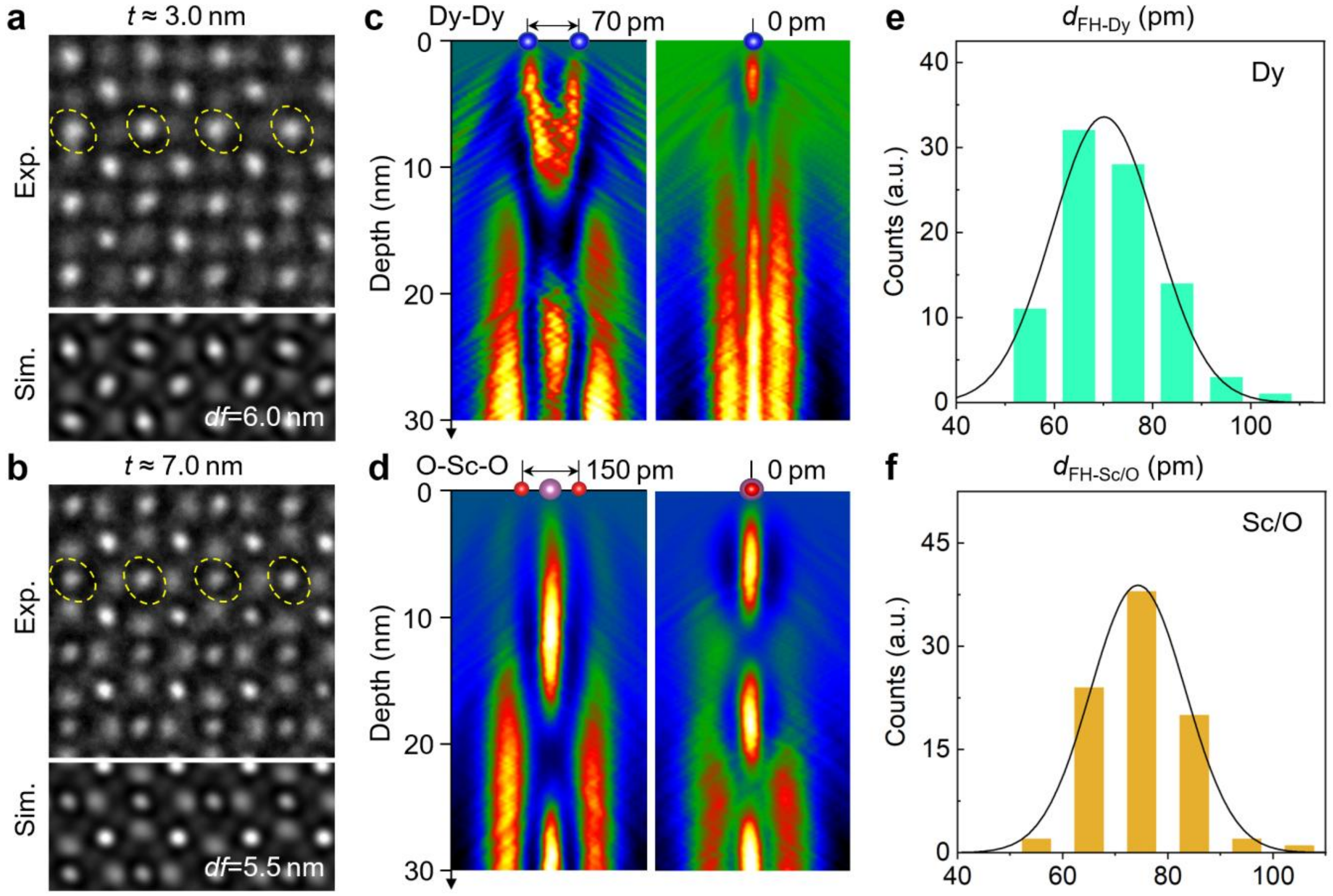

**Supplementary Fig. 2. Depth-dependent structural motifs of $DyScO_3$. a**,**b**, Atomic-resolution Cs-TEM images of thin- and thick-region $DyScO_3$ taken from Fig. 1b (top row) and their best-fit simulated image (bottom row), respectively. **c**,**d**, Depth-dependent electron propagation pathway maps for Dy-Dy and Sc/O columns ($dz = 2\Delta dz = 2a$, $a = 0.4$ nm), respectively. The projection distances for Dy and Sc/O column are 70, 0 pm and 150, 0 pm, respectively. **e**,**f**, Histograms of FWHM values ($d_{FH}$) for Dy and Sc/O atom columns extracted from Fig. 1b, respectively. Their maximum probability values are $d_{FH\text{-}Dy\text{-}MP} = 70.2$ pm and $d_{FH\text{-}Sc/O\text{-}MP} = 74.3$ pm.

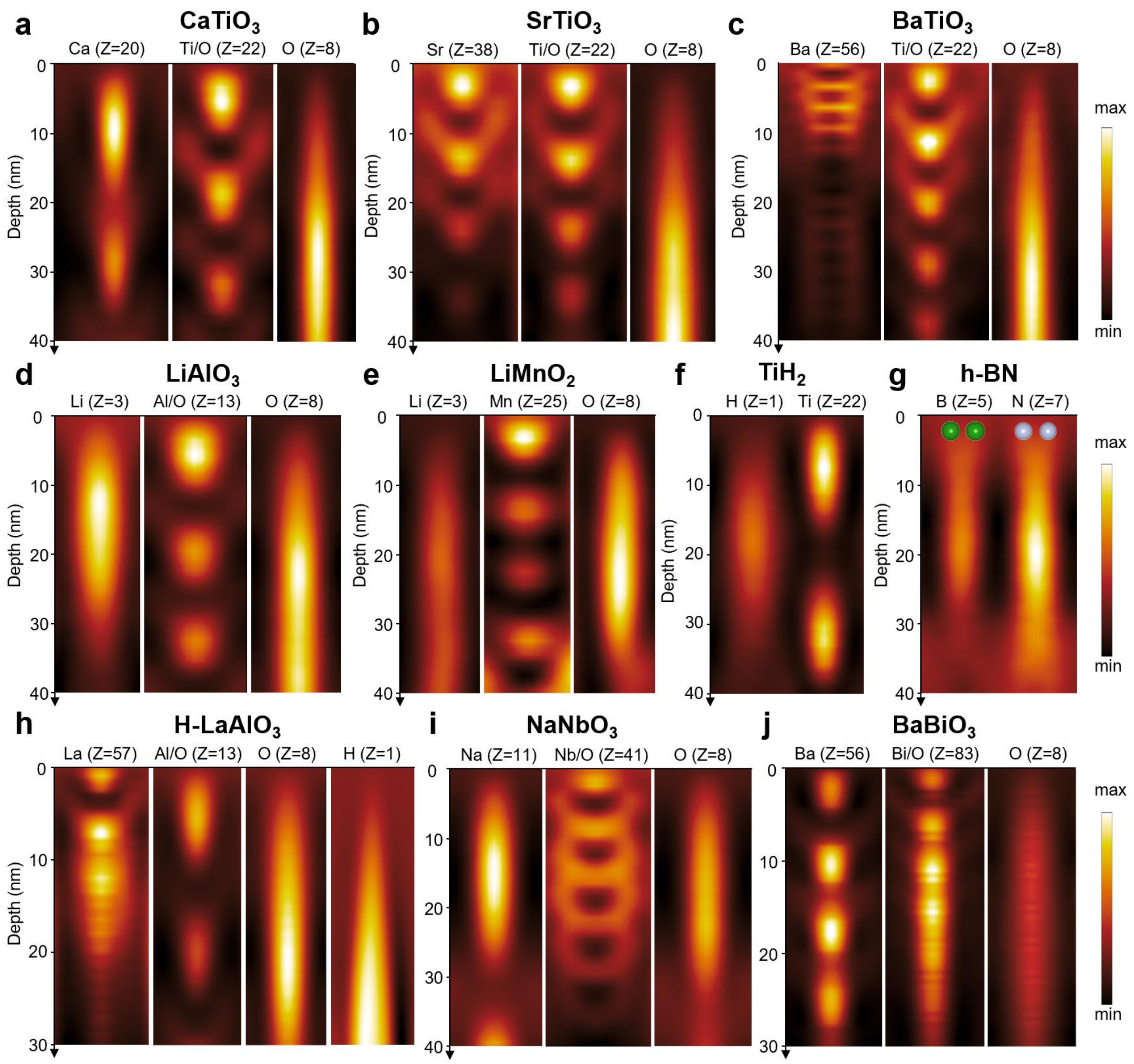

**Supplementary Fig. 3. Interference patterns of channeling electrons for elements in different compounds.** The multi-slice image simulations are performed at Cs = df = 0 nm. **a-c,** $A\mathrm{TiO_3}$ with $A$ = Ca, Sr and Ba. **d**,**e**, $LiAlO_3$ and $LiMnO_2$. **f**,**g**, $TiH_2$ and h-BN. **h-j**, H-$LaAlO_3$, $NaNbO_3$ and $BaBiO_3$. The atomic number Z is labelled near each of the element.

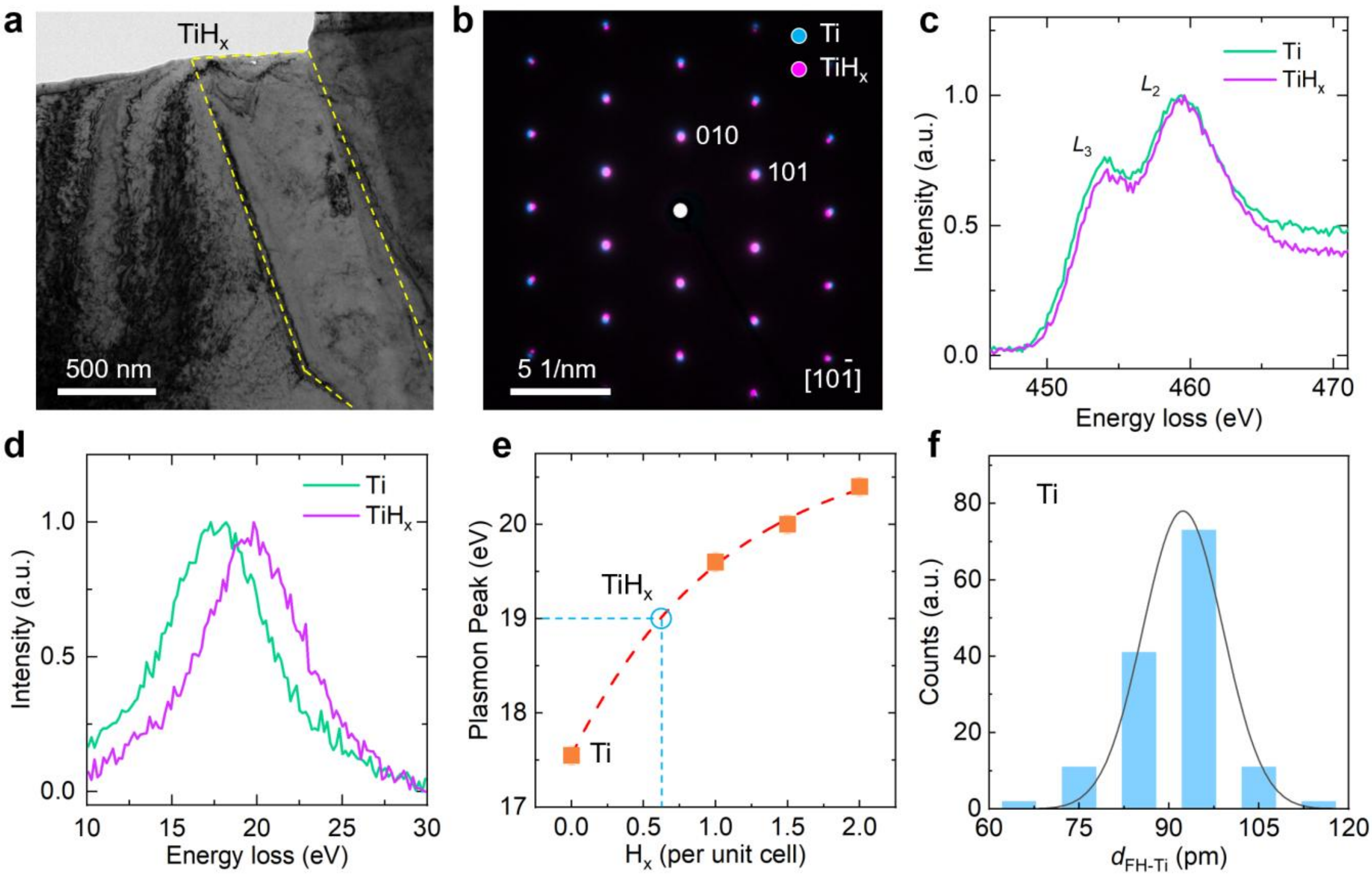

**Supplementary Fig. 4. Crystal and electronic structures of Ti and TiH$_x$. a**, Bright-field TEM image showing morphology of the TiH$_x$ sample, the atomic-resolution image is taken from the edge of a needle-shaped region (marked by yellow dashed lines). **b**, Selected area electron diffraction patterns of Ti and TiH$_x$ taken along [10-1] direction. **c**, Comparison of Ti-$L_{2,3}$ edges between Ti metal and TiH$_x$. **d**, Comparison of bulk plasmon peaks of Ti and TiH$_x$ with their maxima locating at 17.6 eV and 19.0 eV, respectively. Relative to Ti, the peak in TiH$_x$ collected from a near-edge region exhibits a blue shift by 1.4 eV. **e**, Correlation of plasmon peak position with hydrogen concentration among α-Ti, TiH$_x$, γ-TiH, δ-TiH$_{1.5}$, ε-TiH$_2$ according to previous studies[11, 12]. The hydrogen content in TiH$_x$ is determined as $x \approx 0.6$. **f**, Histogram of $d_{FH}$ for Ti -atom columns extracted from Fig. 2e with the MP value at $d_{FH\text{-}Ti\text{-}MP}$ = 93.3 pm. The distribution of $d_{FH\text{-}Ti}$ (Fig. 2d) in Ti metal is very similar to that in TiH$_x$.

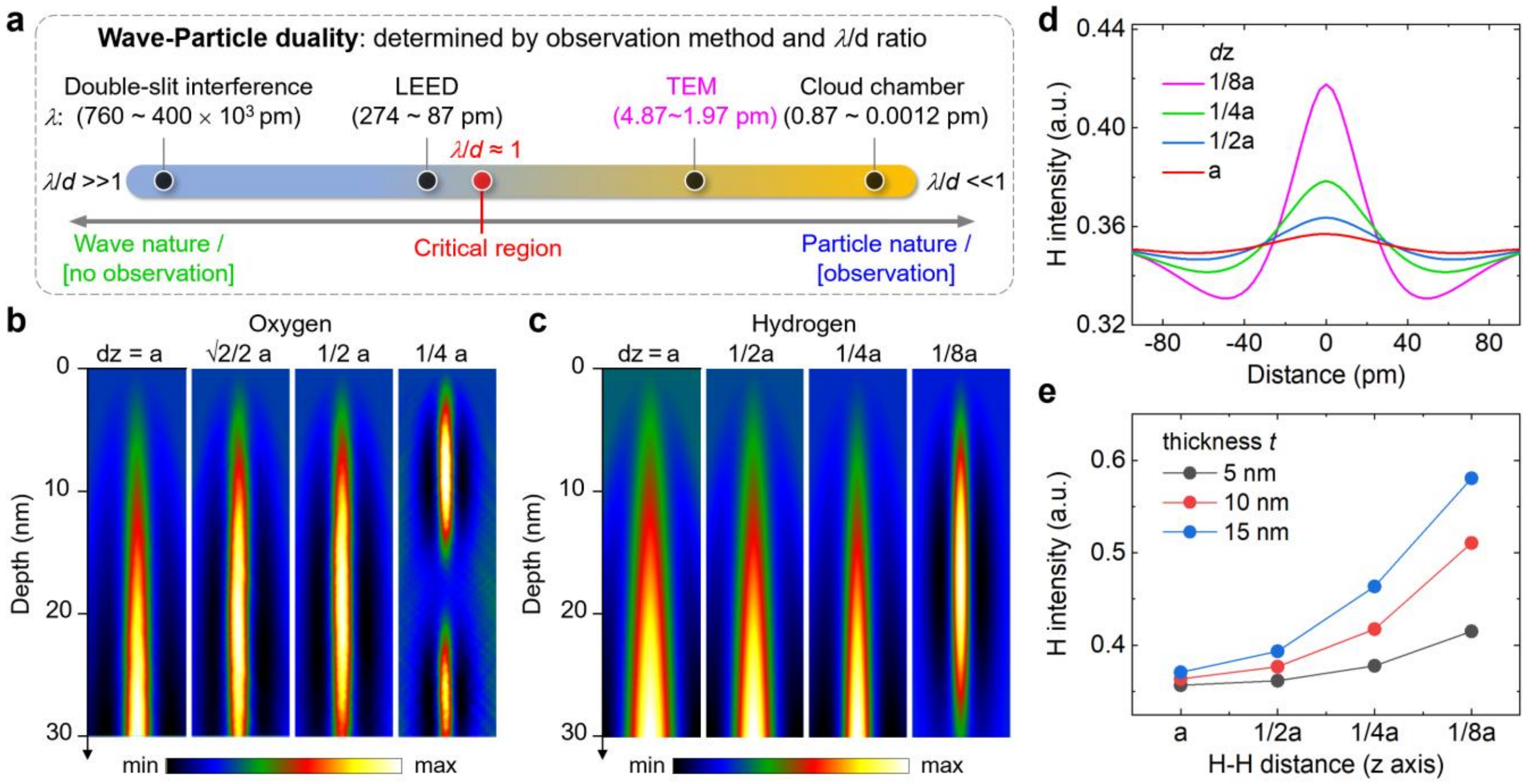

**Supplementary Fig. 5. Ballistic electron propagation-driven SRI of light elements. a**, Relationship of $\lambda$/d ratio with the wave and particle nature in the de Broglie's wave-matter hypothesis. For TEM with $\lambda/d \ll 1$, the particle nature of electrons cannot be ignored. **b**,**c**, Ballistic electron propagation along oxygen and hydrogen column with z-direction spacing from $a$ to $a$/4 and $a$/8, respectively. **d**, Intensity profiles of H column with different H-H spacing along z axis at $t$ = 5 nm. **e**, Peak intensity variation of H columns as a function of H-H spacing along z axis and at different specimen thickness. Similar intensity variation is also observed in the O columns.

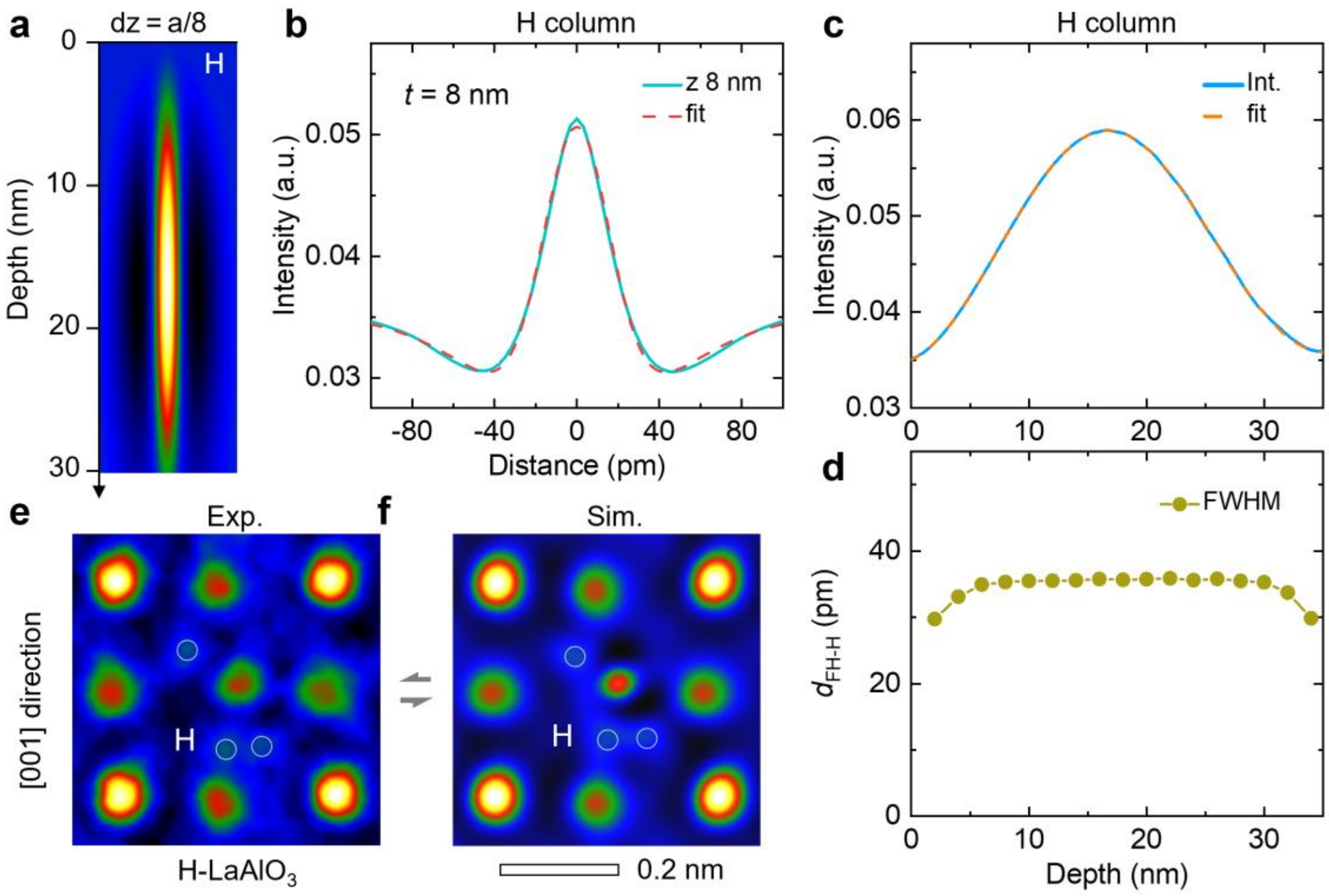

**Supplementary Fig. 6. Quantitative description of ballistic electron propagation along H column.** **a**, The propagation path map of H with $dz = a/8$ ($a$ = 0.4 nm). **b**, Representative cross-sectional intensity profile of (**a**) at $t$ = 8 nm and its fitting by using $I(z) = a_1 \cdot \exp[-(z-z_0)^2/(2\sigma_1^2)] - a_2 \cdot \exp[-(z-z_0)^2/(2\sigma_2^2)] + c$. According to the $d_{FH} = 2\sqrt{(2\ln2)}\sigma$ relation, the inner and outer FWHM values pertinent to $\sigma_1$ and $\sigma_2$ are $d_{FH-H}$ = 38.2 pm and 106.9 pm, respectively. **c**,**d**, The peak intensity $I_{peak}(z)$ and its fitting by Equation (2) as well as $d_{FH-H}(z_i)$ profiles of (**a**) as a function of $z$-direction depth, respectively. The situation for O column is very similar. With $dz = a/4$, the fitting results are $B/A \approx 0.033$, $d_{FH-O} \approx 34$ pm and $\xi \approx 18$ nm. **e**,**f**, Experimental and simulated H-$LaAlO_3$ unit-cell ($t$ = 4.5 nm) structural images, respectively. The latter is obtained by combining wave- ($ABO_3$) and particle-based (H) image simulations.

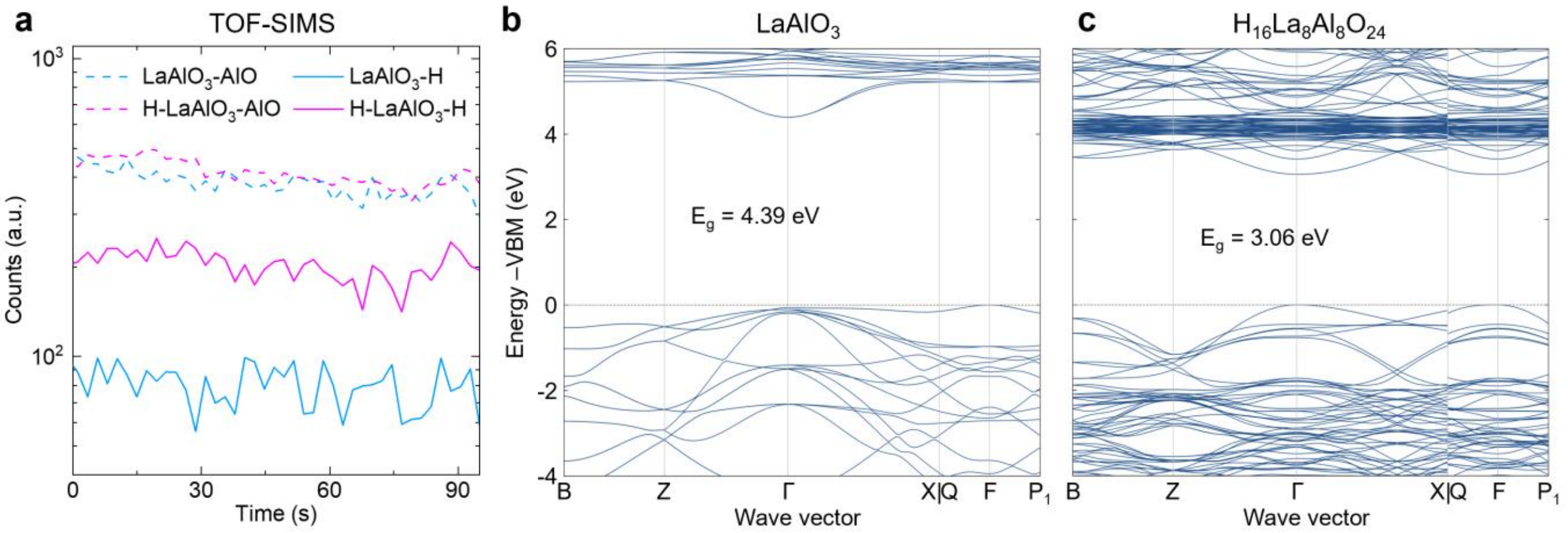

**Supplementary Fig. 7. H distribution and H-induced band structure changes. a**, Comparison of H content between $LaAlO_3$ and H-$LaAlO_3$ crystals obtained by time-of-flight secondary ion mass spectroscopy by taking the AlO content as a reference. **b**,**c**, First-principles calculated band structures of $LaAlO_3$ and $H_{16}La_8Al_8O_{24}$, respectively. Due to hydrogen injection, the band gap value of H-$LaAlO_3$ is reduced by ~1.33 eV.

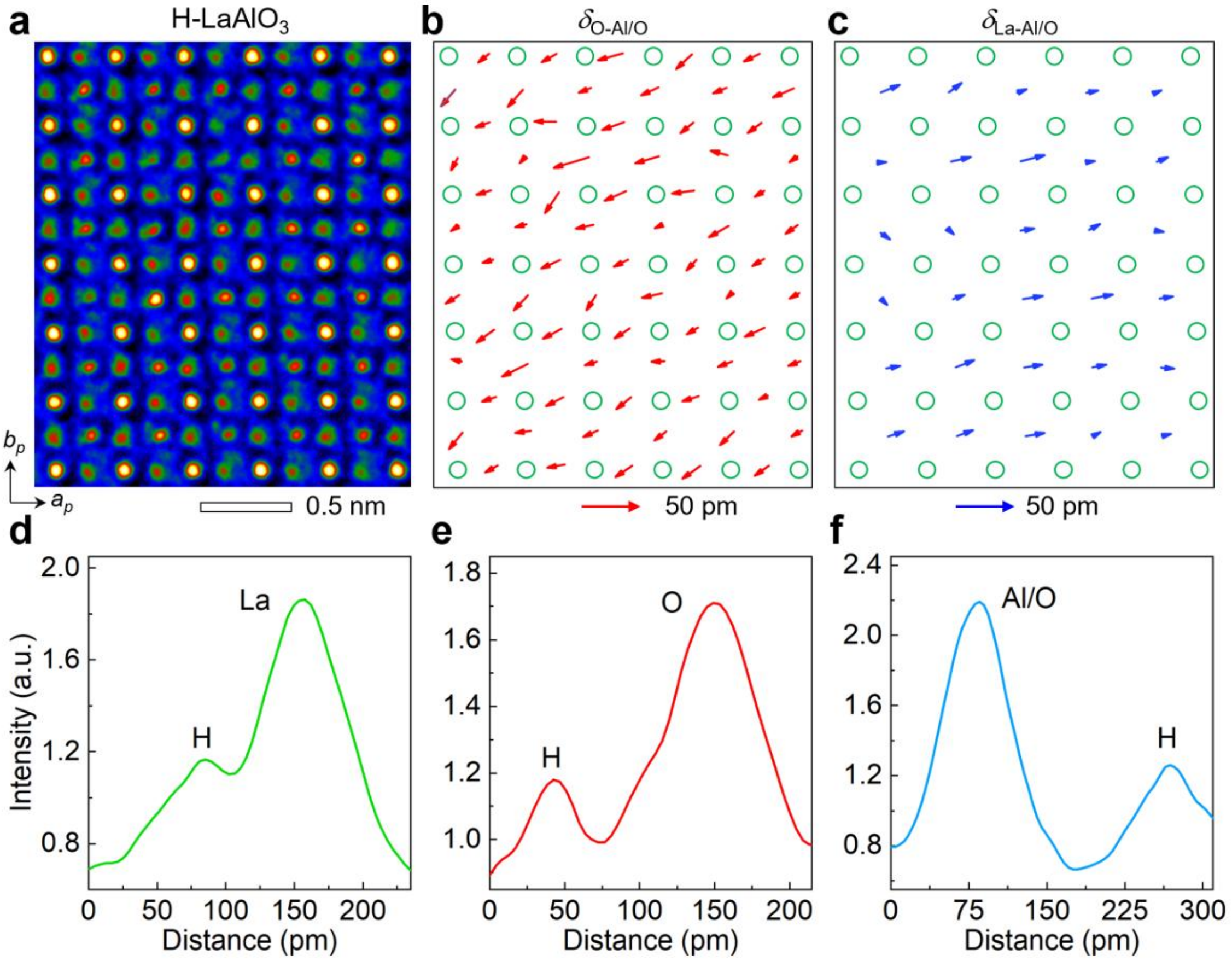

**Supplementary Fig. 8. Polar atomic displacements and H contrast in H-$LaAlO_3$. a**, Atomic-resolution Cs-TEM image taken at $t_0$ = 0 s. **b**,**c**, Relative displacement map of oxygen and La columns with respect to the neighboring Al/O columns, respectively. **d-f**, Representative La-H, O-H and Al/O-H intensity profiles showing the robust and high effective contrast of H with $R_{H/La} \approx 19.3\%$, $R_{H/O} \approx 25.3\%$ and $R_{H/AlO} \approx 21.7\%$, respectively.

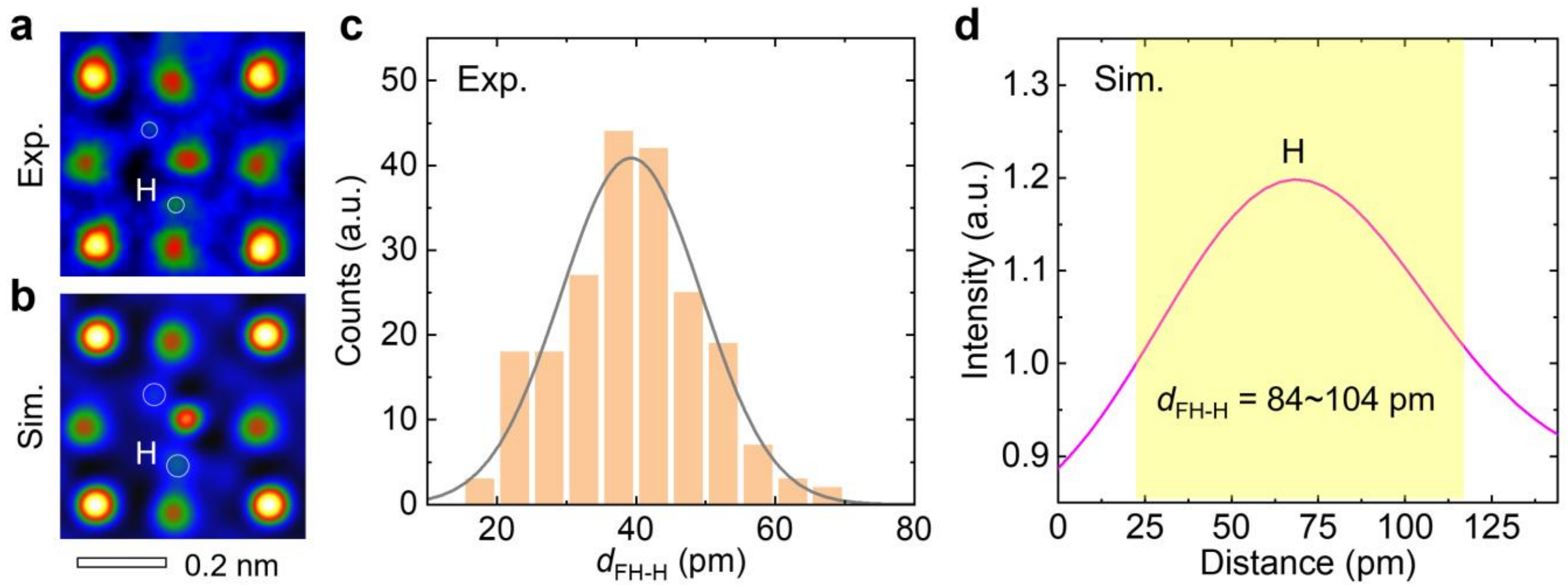

**Supplementary Fig. 9. Comparison of experimental and simulated images of H-$LaAlO_3$ and their $d_{FH-H}$.** **a**,**b**, Experimental with best-fit simulated images of H-$LaAlO_3$ with annotation of interstitial H positions (solid circles), respectively. **c**, Statistical histogram of FWHM values of H ($d_{FH-H}$) measured from the experimental image. A Gaussian peak fitting yields that the MP value locates at $d_{FH-H-MP}$ = 39.3 pm. **d**, Intensity profile of H column in (**b**) showing the large $d_{FH-H}$ value in the simulated image.

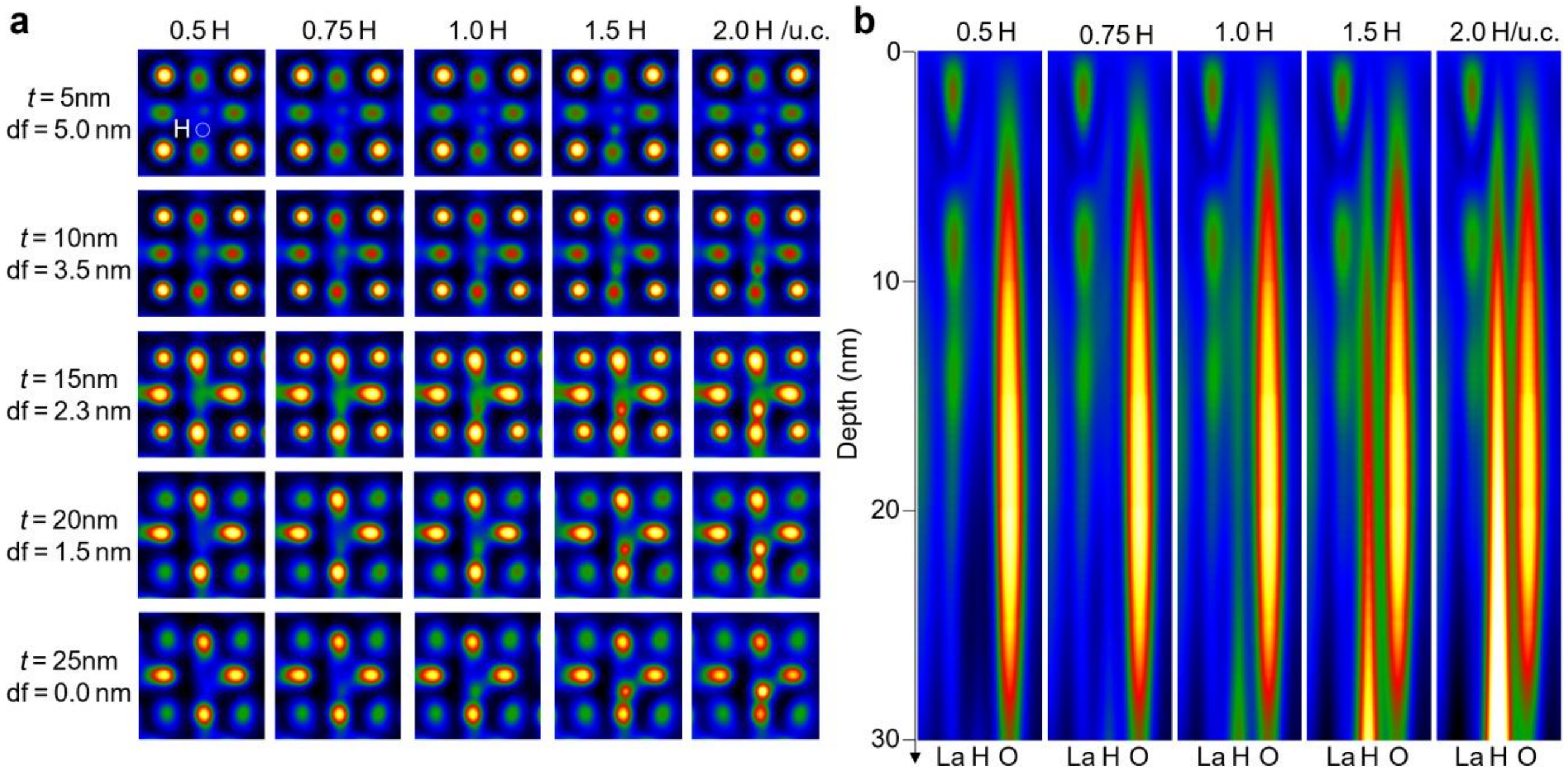

**Supplementary Fig. 10. Thickness dependent H-contrast evolution in H-LaAlO₃.** **a**, Simulated images of La-centered $H_x$-$LaAlO_3$ unit cells (x = 0.5, 0.75, 1.0, 1.5 and 2.0 H per unit cell) by combining wave-based multi-slice simulations of $LaAlO_3$ lattices (*Cs* = -5.5 μm) and particle-based H column at different thickness *t* and defocus *df* values. **b**, Depth dependent La-H-O-column intensity maps with different x values, with the signal extracted from images simulated at *df* = 3.5 nm. The upper thickness limit of structural images appears to be above ~ 25 nm.

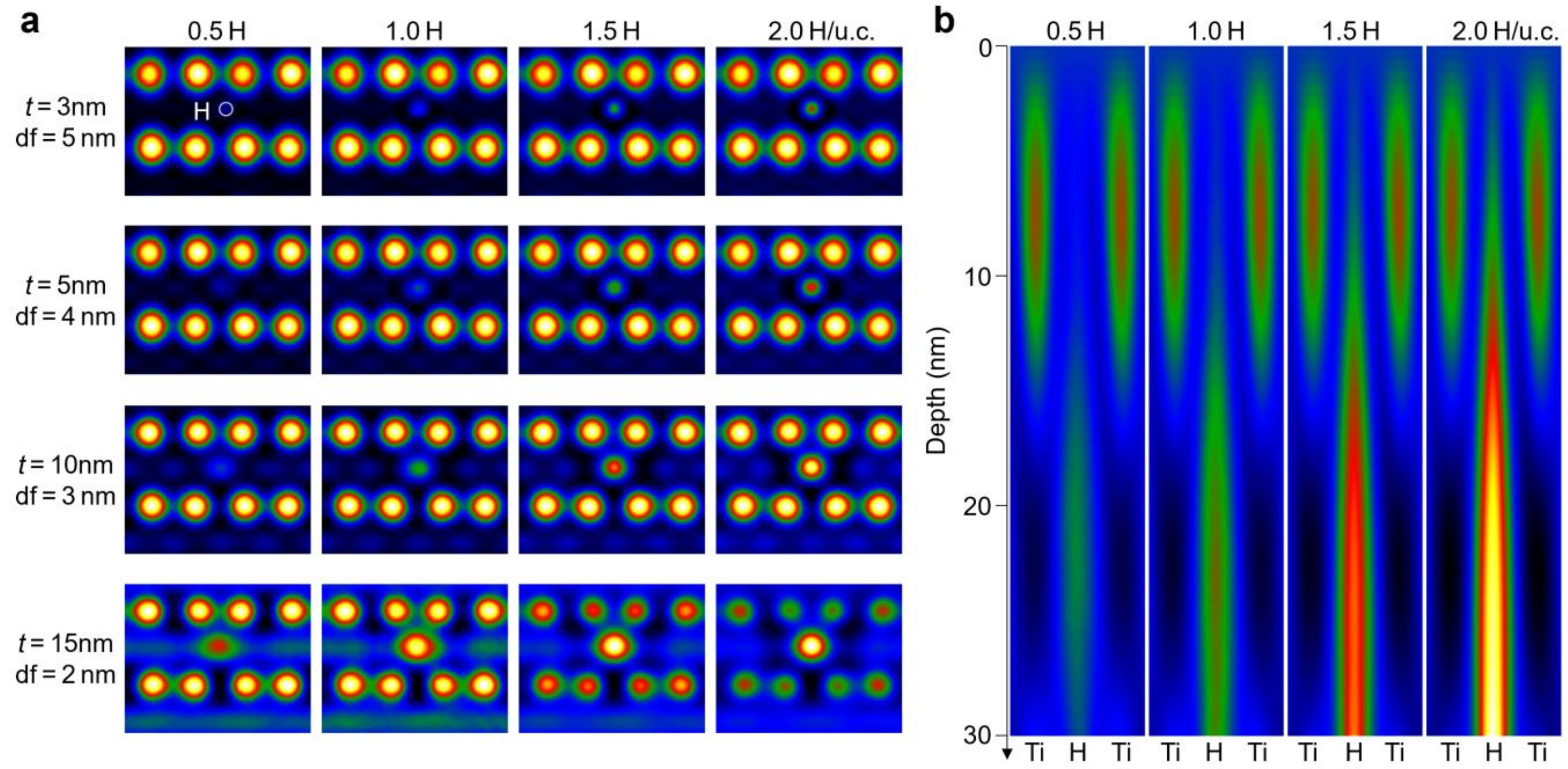

**Supplementary Fig. 11. Thickness dependent H-contrast evolution in $TiH_x$. a**, Simulated images of $TiH_x$ (x = 0.5, 1.0, 1.5 and 2.0 H per unit cell) by combining wave-based multi-slice simulations of Ti lattices (*Cs* = -10 μm) and particle-based H column at different thickness *t* and defocus *df* values. **b**, Depth dependent Ti-H-column intensity maps with different x values, with the signal extracted from images simulated at *df* = 5 nm. The upper thickness limit of structural images is ~ 15 nm.

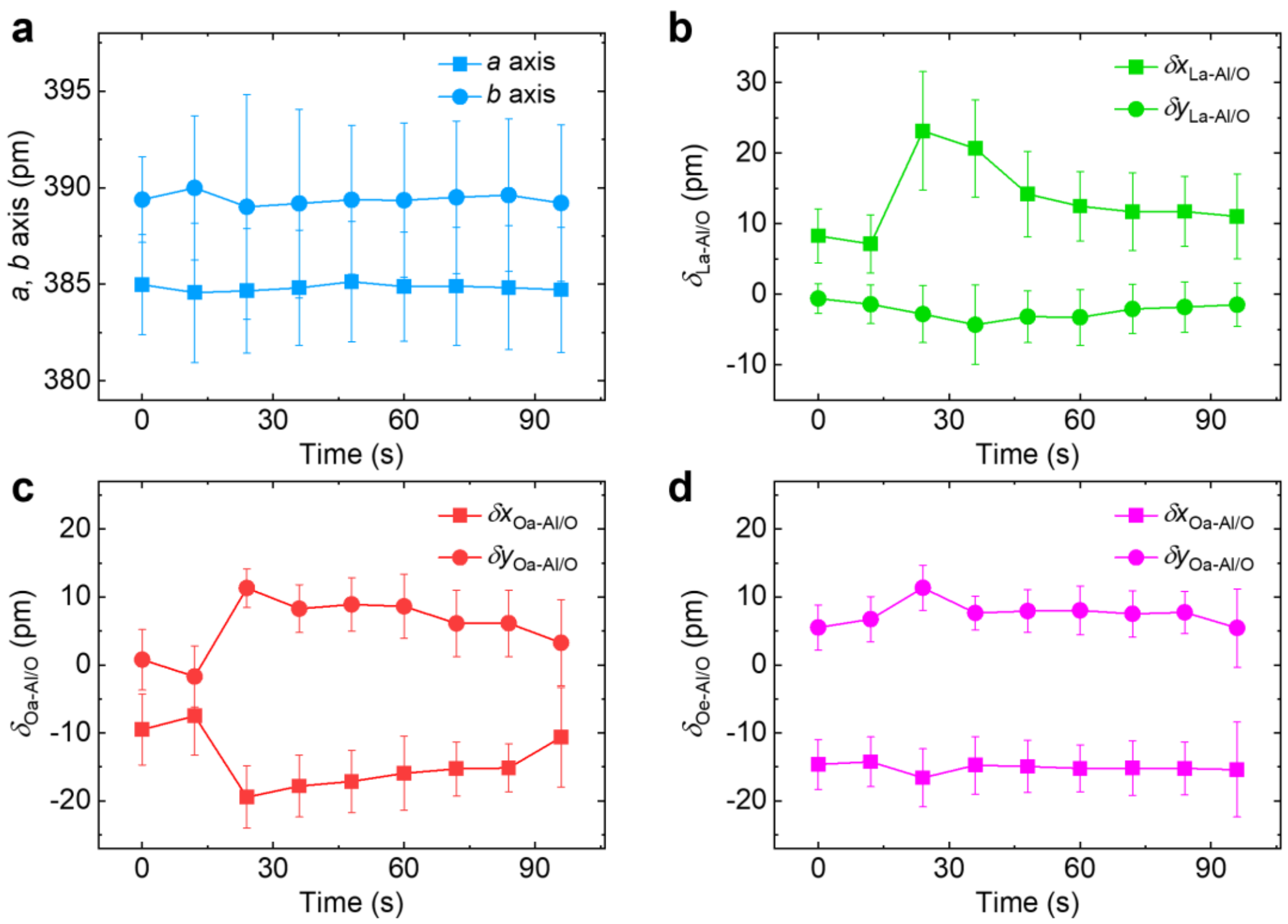

**Supplementary Fig. S12. Temporal resolution lattice parameter and atomic displacement in H-LaAlO$_3$.** **a**, Averaged lattice parameter change. **b**,**c**, Abrupt La-atom and Oa-atom shifts driven by collective hydrogen tunneling around $t_0 = 24$ s, respectively. **d**, Time dependent Oe-atom shifts.

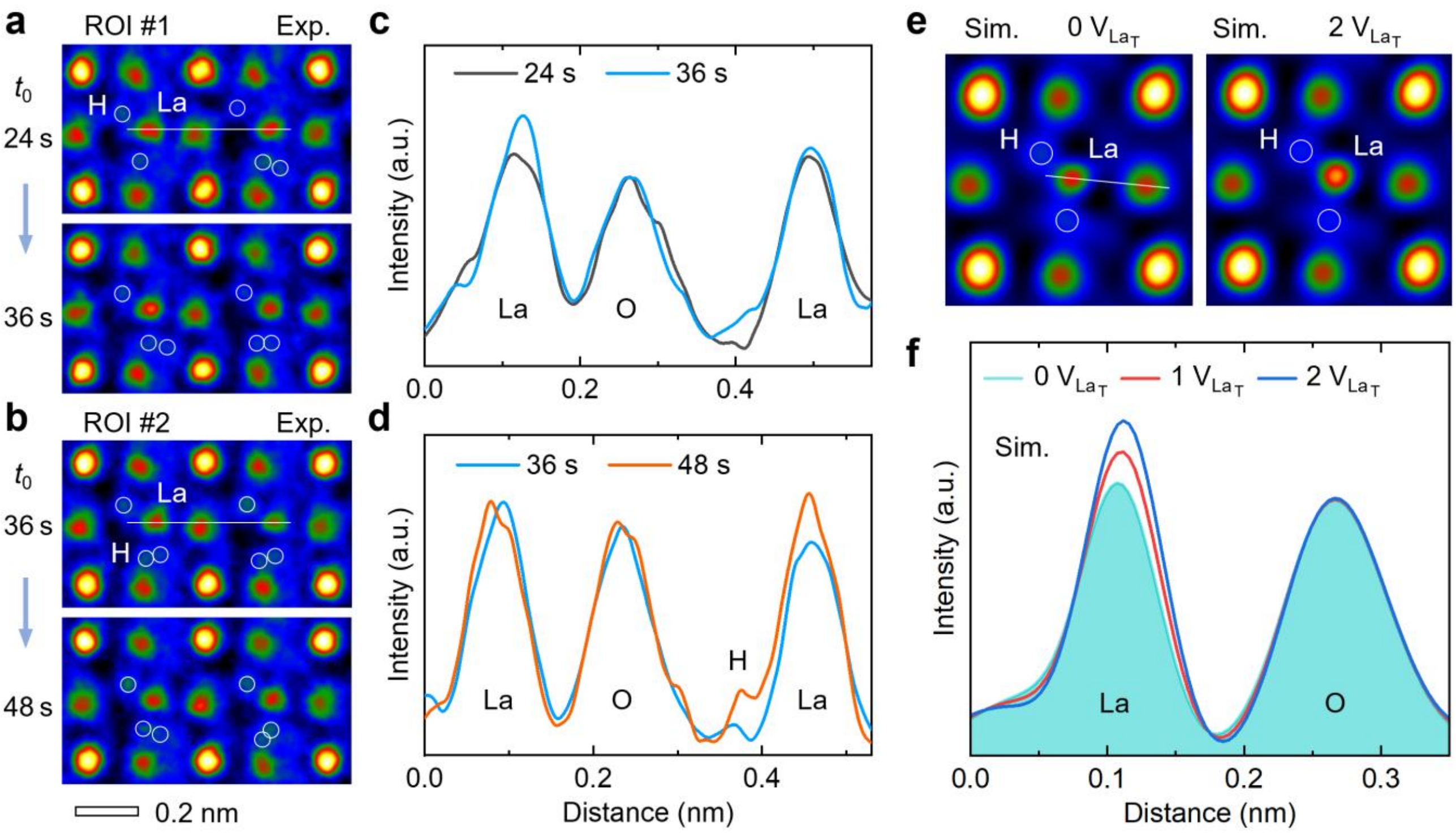

**Supplementary Fig. S13. Hydrogen-induced La-atom stripping in H-LaAlO$_3$. a**,**b**, Time-series Cs-TEM images for two region of interest (ROI). The white circle marks the hydrogen atom. **c**,**d**, Intensity profiles extracted separately from the La-O-La columns marked by a white line in (**a**) and (**b**). **e**, Simulated Cs-TEM images on the structure without and with two La-atom vacancies ($V_{La}$) on the sample surface. Each H site marked by white circles has 0.5 H per unit cell. **f**, Intensity profiles of La-O columns without and with removal of one and two La atoms from top (T) surface of the structure. The position of extracting the intensity profile is marked by a white line in (**e**).

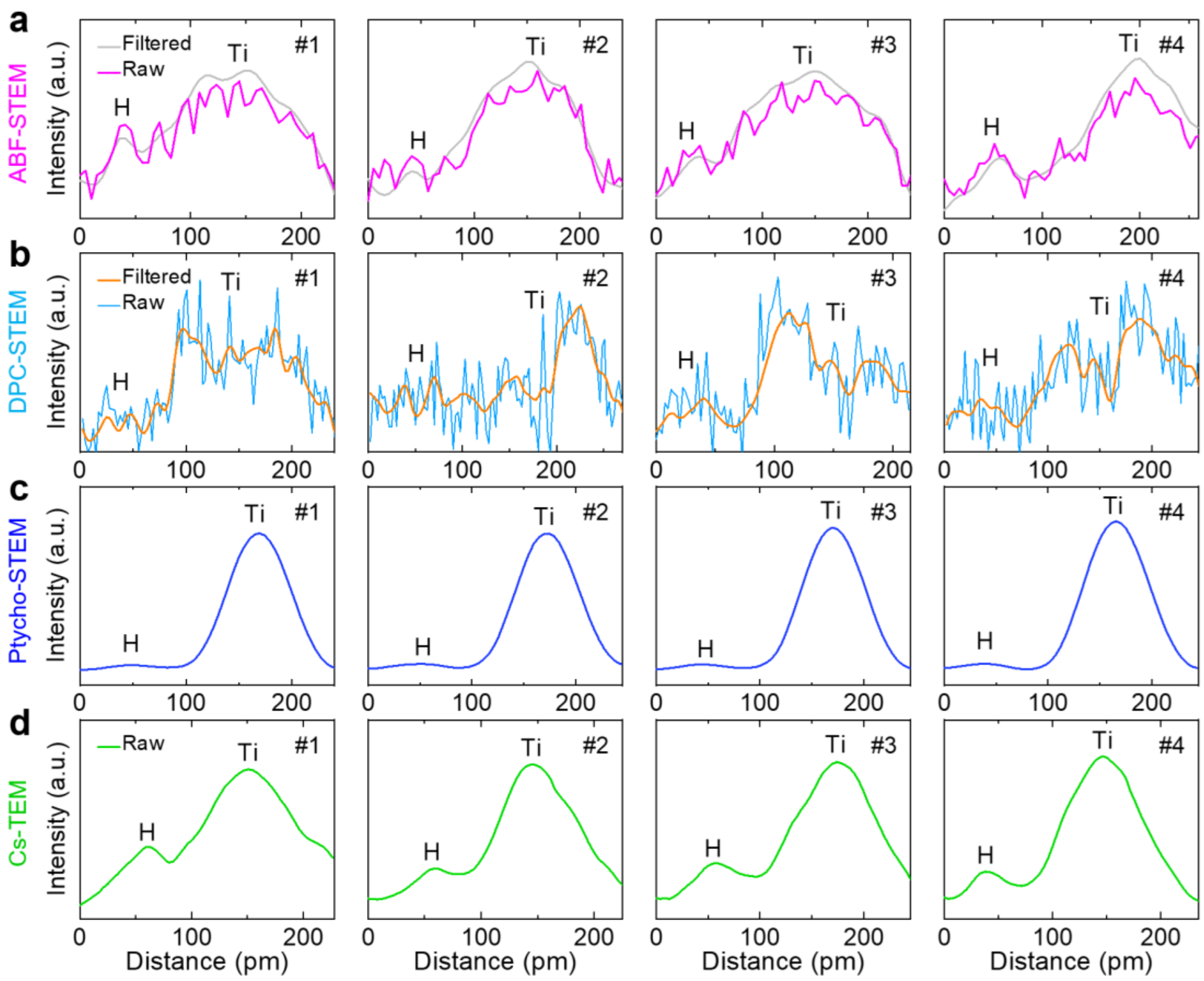

**Supplementary Fig. 14. Comparison of Ti-H column intensity profiles in $TiH_x$. a-d**, Representative intensity profiles, raw and nonlinear filtered ones, extracted from the ABF-, DPC-, ptychography-STEM and Cs-TEM images shown in **Table 1**, respectively.

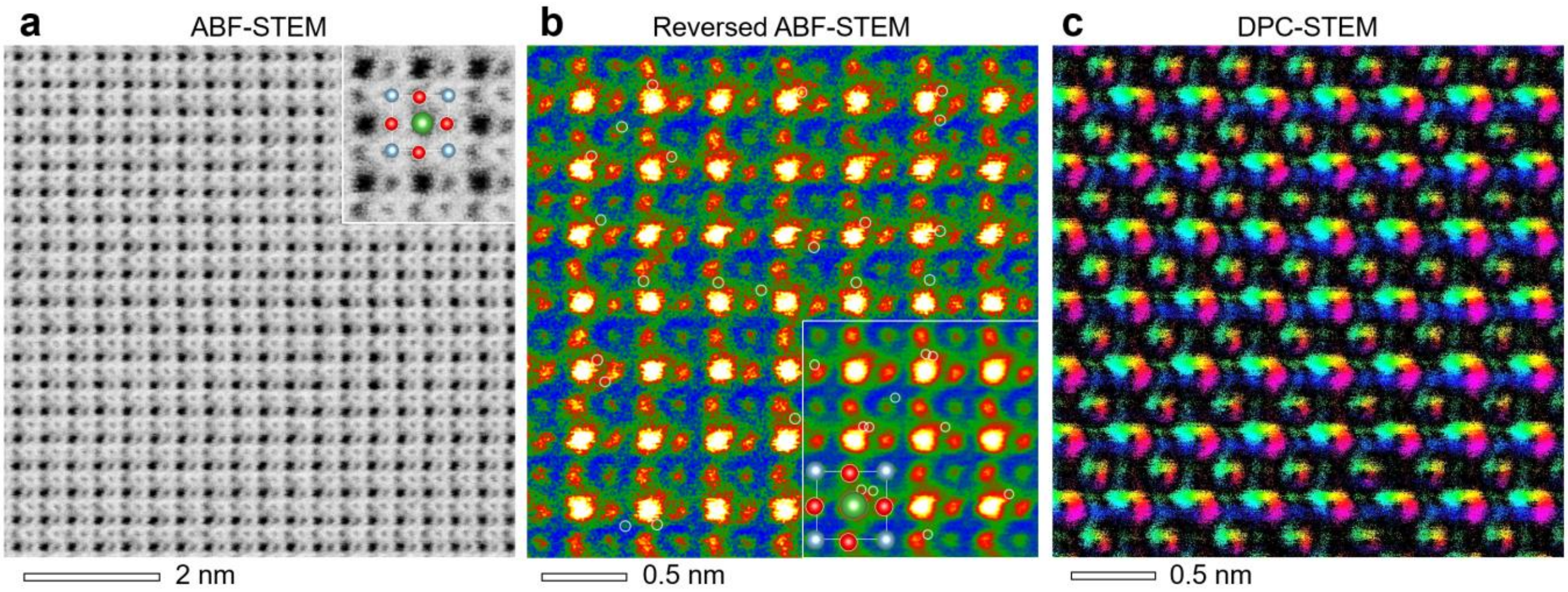

**Supplementary Fig. 15. ABF- and DPC-STEM results of H-LaAlO$_3$. a**, Original and contrast-reversed atomic-resolution STEM-ABF images (La-green, Al/O-blue, O-red) collected along [001] direction. **b**, An enlarged view of the ABF image with false color and ABS (average background subtraction) filtered one (inset). **c**, Corresponding atomic-resolution DPC-STEM image with nearly absent H-atom contrast.

## Supplementary References